\documentclass{elsart}

 \usepackage{epsfig}

\usepackage{amssymb}
\begin{document}

\begin{frontmatter}



\title{Dynamics of Defects in the Vector Complex
Ginzburg-Landau Equation}


\author[imedea,mardp]{Miguel Hoyuelos}
\author[imedea]{, Emilio Hern\'andez-Garc\'{\i}a}
\author[imedea]{, Pere Colet}
\author[imedea]{and Maxi San Miguel}

\address[imedea]{Instituto Mediterr\'aneo de Estudios Avanzados, IMEDEA
\thanksref{www}
(CSIC-UIB),\\ Campus Universitat Illes Balears, E-07071 Palma de
Mallorca, Spain.}
\thanks[www]{\tt http://www.imedea.uib.es/PhysDept/}

\address[mardp]{Departamento de F\'{\i}sica, Facultad de Ciencias Exactas y
Naturales, Universidad Nacional de Mar del Plata, Funes 3350, 7600
Mar del Plata, Argentina}

\date{28 September 2001}

\begin{abstract}
Coupled Ginzburg-Landau equations appear in a variety of contexts
involving instabilities in oscillatory media. When the relevant
unstable mode is of vectorial character (a common situation in
nonlinear optics), the pair of coupled equations has special
symmetries and can be written as a {\sl vector complex
Ginzburg-Landau equation}. Dynamical properties of localized
structures of topological character in this vector-field case are
considered. Creation and annihilation processes of different kinds
of vector defects are described, and some of them interpreted in
theoretical terms. A transition between different regimes of
spatiotemporal dynamics is described.
\end{abstract}


\end{frontmatter}

\section{Introduction}
\label{sect:intro}

Localized structures, objects with some kind of particle-like
behaviour, can be found in a variety of non-linearly evolving
fields.  Examples are vortices in fluids, superfluids and
superconductors, propagating pulses of excitation in nerve
systems, solitary waves in chemical media, in parametrically
driven surfaces, and in granular media, among others
\cite{riecke-pismen}. An understanding of complex evolving
configurations can be sometimes achieved in terms of the
interaction rules of the particle-like entities.  In most cases,
however, numerical integration of the evolution equations is the
most powerful tool to investigate the role of the localized
structures.

Nonlinear optical cavities have been specially prolific in
providing examples of localized structures \cite{nonlinearoptics}.
They appear in the tranverse profile of the field and can take the
form of vortices, or of bright or dark dissipative spatial
solitons. It is usually considered that the polarization degree of
freedom of the electromagnetic field is fixed either by material
anisotropies or by experimental arrangement. Thus the description
of the dynamics is done in terms of a scalar field. However in the
cases in which the polarization of the light is not fixed, the
vector nature of the electromagnetic field leads to striking
topological phenomena \cite{bhandari}. Recently reported examples
of localized structures for which the polarization plays a
fundamental role include points of zero amplitude in both
components of the field in a periodic elliptically polarized
background, found in type II OPO \cite{marco} and dots of low
amplitude in a circularly polarized component of the field in an
almost circularly polarized background found in self-defocusing
vectorial Kerr resonators \cite{rafa}.

The scalar Complex Ginzburg-Landau (CGL) equation is considered a
para\-digm model for the qualitative description of general
non-linear oscillatory media (see, for example,
\cite{cross,bohr,aranson_kramer}). The vector complex
Ginzburg-Landau (VCGL) equation
\cite{gil,gilIJBC,pismenPRL,pismen,sanmiguel} plays the same role
when the order parameter is of vector character, as is the case
for the electromagnetic field when the polarization is not fixed.
For example, the VCGL equation is an appropriate model for laser
emission from wide-aperture resonators close to the lasing
threshold \cite{sanmiguel} in the absence of
polarization-selecting cavity features. Different kinds of
localized structures
\cite{gil,gilIJBC,pismenPRL,pismen,hernandez,hoyuelos,prl} are
present in the dynamic states of the twodimensional VCGL equation.
From a topological point of view, they are {\sl defects}, that is,
places where the field state departs from some basic ordered
state. These objects carry topological properties which endorse
them with a characteristic stability and robustness.

In this Paper we study the dynamical properties of these defects
for several dynamical regimes of the VCGL equation. In addition to
reviewing and adding details to results previously presented,
which address the synchronization properties of spatiotemporally
chaotic states \cite{hernandez}, the identification of a
transition from a {\sl glass} to a {\sl gas} phase
\cite{hoyuelos}, and the formation and annihilation processes
leading to the different types of defects \cite{prl}, we interpret
some of these findings in terms of the stability properties of the
waves emitted by the defect cores, and in terms of a recent
perturbative analytic approach \cite{aranson-pismen}. Since we are
interested mostly in the dynamics of topological defects, we
consider here just the twodimensional VCGL case, for which point
defects are topologically stable. For VCGL dynamics in one spatial
dimension we refer the reader to
\cite{amengual,hernandez,montagnePL}.

Section \ref{sect:eqs} presents the basic equations and
properties. The different topological defects are described in
Sect.~\ref{sect:defects}, and Sect.~\ref{sect:dynamics} is devoted
to their dynamical properties. A description of a {\sl gas-like}
phase is presented in Sect.~\ref{sect:gas}.  In
Sect.~\ref{sect:gamma1} we briefly discuss a situation of phase
separation. Finally (Sect.~\ref{sect:conclusions}) we summarize
our conclusions. The Appendix addresses the stability of waves
emitted by vector defects.

\section{The vector Complex Ginzburg-Landau equation}
\label{sect:eqs}

Extended systems close to a Hopf bifurcation at zero wavenumber
are described by the CGL equation :
\begin{equation}
\partial_t A = A + (1 + i\alpha) \nabla^2 A - (1 + i\beta)|A|^2 A.
    \label{cgle}
\end{equation}
For systems where the relevant unstable mode is of vectorial
character, Eq. (\ref{cgle}) is generalized to the VCGL equation,
\begin{equation}
\partial_t {{\bf A}} = {\bf A} + (1 + i\alpha) \nabla^2 {\bf A} - (1 + i\beta)
\left[({\bf A} \cdot {\bf A}^*) {\bf A} + \frac{(\gamma-1)}{2} ({\bf A}
\cdot {\bf A}) {\bf A^*} \right].
    \label{vcglexy}
\end{equation}
In the simplest case $\bf{A}$ has two complex components. This is
precisely the case of optical systems, for which, ${\bf A}= (A_x,
A_y)$ describe the complex slowly varying amplitude of the
electric field, where $A_x$ and $A_y$ are the cartesian
components. The right and left circularly polarized components
$(A_+,A_-)$ are related to them via the relations $A_x = (A_+ +
A_-)/\sqrt{2}$ and $A_y = (A_+ - A_-)/(i\sqrt{2})$. In terms of
the circular components, Eq. (\ref{vcglexy}) reads
\begin{equation}
\partial_t A_\pm = A_\pm + (1 + i\alpha) \nabla^2 A_\pm - (1 + i\beta)
(|A_\pm|^2 + \gamma |A_\mp|^2) A_\pm.
    \label{vcgle}
\end{equation}
 The coefficient $\gamma$, which
in general can be complex, gives the coupling
between the components. For $\gamma = 0$ we recover a
pair of uncoupled equations of the form (\ref{cgle}). We will
often refer to Eq.~(\ref{cgle}) as the {\sl scalar} CGL, to
contrast it with Eqs.~(\ref{vcglexy}) or (\ref{vcgle}). For
$\gamma\neq 0$, Eq.~(\ref{vcgle}) can be thought as a particularly
symmetric example (for example group velocity terms are absent) of
a pair of coupled complex Ginzburg-Landau equations of the kind
usually encountered in wave competition situations \cite{cross}.

When Eq.~(\ref{vcglexy}) is derived for the Hopf bifurcation
leading to laser emission\cite{sanmiguel}, $\alpha$ is related to
the strength of diffraction, and $\beta$ to the nonlinear
frequency detuning. In this case, the so called Benjamin-Feir
stability criterion $1+\alpha\beta > 0$ is always satisfied, which
means that there are always some plane waves which are stable
against long wavelength perturbations. In addition, for laser
systems, $\gamma$ turns out to be a real parameter, and
$\gamma>-1$. In this Paper, we will restrict our study to these
ranges of parameters.

Equation (\ref{vcgle}) has a family of plane-wave solutions of the
form:
\begin{equation}
A_\pm = Q_\pm e^{-i({\bf k_\pm} . {\bf r} - \omega_\pm t +
\phi_{0\pm})} \ \ .
    \label{plana1}
\end{equation}
Within this family, the simplest solutions are the circularly
polarized traveling waves, for which one of the components
is identically zero:
\begin{equation}
Q_+=1-k^2, Q_-=0, \omega_+= -\beta + (\beta - \alpha)k^2
\label{circular+}
\end{equation}
or
\begin{equation}
Q_+=0, Q_-=1-k^2, \omega_-= -\beta + (\beta - \alpha)k^2
\label{circular-}
\end{equation}
Within the family (\ref{plana1}) there are also solutions in
which both components coexist.
For wavevectors with the same modulus, $k_+=k_-=k$, we have
\begin{eqnarray}
 Q_+^2 &=& Q_-^2 = Q^2 =\frac{1-k^2}{1+\gamma} \nonumber \\
 \omega_+ &=& \omega_- = \omega = -\beta + (\beta - \alpha)k^2 .
    \label{omega}
\end{eqnarray}
We can have a linearly polarized traveling wave (${\bf k_+} = {\bf
k_-}$)  or a standing wave with periodic linear polarization
(${\bf k_+} = -{\bf k_-}$).  It is also possible to have
depolarized solutions \cite{sanmiguel} with $k_+ \neq k_-$:
\begin{eqnarray}
Q_\pm &=& (1-\gamma +
\gamma k_\pm^2 - k_\mp^2)/(1-\gamma^2) \nonumber \\
\omega_\pm &=& -\alpha k_\pm^2 -
\beta(Q_\pm^2 + \gamma Q_\mp^2)
\label{depolarized}
\end{eqnarray}

The qualitative behavior for the VCGL equation (\ref{vcgle}) for
$|\gamma|<1$ is rather different from the one for $\gamma>1$. In
the second case the nonlinear competition between $A_+$ and $A_-$
tends to favor one of them against the other, so that in regions
where $A_+$ is developed, $A_-$ generally vanishes, and viceversa.
There are stable circularly polarized solutions of the form
(\ref{circular+}) or (\ref{circular-}) while solutions of the form
(\ref{omega}) which involve the coexistence of both fields are
linearly unstable. On the contrary, for $|\gamma|<1$, wave
coexistence is generally achieved. Circularly polarized solutions
of the form (\ref{circular+}) or (\ref{circular-}) are linearly
unstable while there are stable solutions of the form
(\ref{omega}) and (\ref{depolarized}). The character of the
topological defects in both cases is different
\cite{pismenPRL,pismen}. In
Sects.~\ref{sect:gamma1}-\ref{sect:gas} we will consider the case
$|\gamma|<1$. The case $\gamma>1$ will be discussed in
Sect.~\ref{sect:gamma1}.

For $\gamma$ real, the point in parameter space $\alpha=\beta=0$
represents a special case of particularly simple dynamics. The
VCGL equation becomes then a potential system, or more precisely
\cite{potential1,potential2}, a `relaxational gradient' dynamical
system. Dynamics at long times approaches steady states that are
identified as the minima of a Lyapunov potential. This is the case
which is considered in greater detail in \cite{pismenPRL,pismen}.
A more general situation is $\alpha=\beta$. In this case the
system can be classified as `relaxational non-gradient'
\cite{potential1,potential2}, and although a Lyapunov functional
can be still identified, the attractor for the dynamics is no
longer steady in general. The limiting `conservative case'
$\alpha=\beta\rightarrow\infty$, also addressed in
\cite{pismenPRL,pismen}, leads to coupled nonlinear
Schr\"{o}dinger (or Gross-Pitaevskii) equations. This last limit
is of great interest for the description of multicomponent Bose
condensates \cite{bose} or the propagation of vector solitons in
optical fibers \cite{fiber}.

\section{Scalar and vector defects for $|\gamma|<1$.}
\label{sect:defects}

Starting from random initial conditions, even if the condition for
existence of stable plane waves ($1+\alpha\beta>0$) is satisfied,
solutions of (\ref{vcgle}) usually do not evolve to a simple plane
wave (\ref{plana1}) due to the presence of defects: points where
the phase of $A_+$ or $A_-$ is undefined and the corresponding
amplitude is equal to zero. Around each defect, a spiral wave
develops (except if $\alpha = \beta$) that, far from the defect
core, approaches a plane wave with a particular wave number which
has been dynamically selected by the presence of the defect. For
example, Fig.~\ref{fig:g01} shows different views of a
configuration obtained after time evolution \cite{numerics}
starting from random initial conditions, for $\alpha=0.2$,
$\beta=2$, and $\gamma=0.1$. The configuration consist of wave
domains separated by shock waves. There is a defect in both
components (the black dot) in the center of each domain.
Inhomogeneities and defects in just one of the components have
been expelled away from the defect core with a given group
velocity.

Therefore, two different kinds of defects are readily identified:
{\sl Vector defects} are points at which the two components $A_+$ and
$A_-$ simultaneously vanish (and thus the same happens for $A_x$ and
$A_y$ at the same point). {\sl Scalar defects} ({\sl mixed defects} in
the nomenclature of
\cite{pismenPRL,pismen}) are those at which only one of the two
fields $A_+$ or $A_-$ vanish. In this case the vector ${\bf A}$ is
not zero, and the components $A_x$ and $A_y$ do not necessarily
vanish, but they have still a singular topological structure to be
described below. The following ansatz describes the
field ${\bf A}$ close to a scalar or to a vector defect:
\begin{equation}
A_\pm(r,\theta) = R_\pm(r) e^{i\phi_\pm (r,\theta)}.
    \label{ansatz}
\end{equation}
Where $(r,\theta)$ are the polar coordinates with the origin
of coordinates at the defect core.
The amplitude of $A_+$ and (or) $A_-$ go to zero at the core of a
vector (scalar) defect: $R_+$ and (or) $R_- \rightarrow 0$ for $r
\rightarrow 0$.  The phase of the component which vanishes at $r=0$
is
\begin{equation}
\phi_i(r,\theta) = n_i \theta + \psi_i(r) - \omega_i t +
\phi_{0i} \ ,
    \label{fase}
\end{equation}
where the subindex $i$ stands for $+$ or $-$. From expression
(\ref{fase}), $n_i$ is seen to satisfy
\begin{equation}
n_i = \frac{1}{2\pi}\oint_{\Gamma} \nabla \phi_i \cdot d{\bf r} ,
\label{charge}
\end{equation}
where $\Gamma$ is a closed path around the defect. This identifies
$n_i$ as the topological charge associated to the singularity in
the field $A_i$. It necessarily takes integer
values. In our numerical simulations we have never found for it a
value different from $0$, $+1$ or $-1$. $\omega_i$ is the rotation
frequency of the spiral wave in $A_i$. For $r \rightarrow \infty$,
$\psi_\pm(r) \rightarrow kr$.

It is possible to make a classification of defects using
topological arguments \cite{pismenPRL,pismen}. Although this is
not the way in which it was originally introduced, the
classification becomes particularly simple in terms of the charges
(\ref{charge}). As stated before, a {\em vector} defect is the one
at which both components of the field vanish. Thus, necessarily
both $n_+$ and $n_-$ are nonvanishing. A vector defect is of type
{\em argument} when the charges in the two field components have
the same signs, i.e., when $n_+ = n_- = 1$ or $n_+ = n_- = -1$. If
the charges are of opposite signs, i.e., when $n_+ = -n_- = 1$ or
$n_+ = -n_- = -1$, the vectorial defect is of {\em director} type.
A {\em scalar} defect is the one at which just one component of
the field, but not the other, vanish. This implies that only one
of the two charges is different from zero: $n_+=
\pm 1$ and $n_- = 0$, or $n_- = \pm 1$ and $n_+ = 0$
(see Fig.~\ref{fig:esquema}).

Numerically we observe that for vectorial defects both fields have
identical modulus near the core: $|A_+| = |A_-|$, thus
$R_+(r)=R_-(r)= R(r)$. Then close to a vector defect, Eq.
(\ref{vcgle}) becomes
\begin{equation}
\partial_t A_{\pm}= A_{\pm} + (1 + i\alpha) \nabla^2 A_{\pm} -
(1 + i\beta)(1+\gamma)|A|_{\pm}^2 A_{\pm}.
    \label{corevector}
\end{equation}
Each component is described by a scalar CGL equation (\ref{cgle})
for a rescaled value of the common amplitude, $\tilde A \equiv
\sqrt{1+\gamma}A_{\pm}$. In the rescaled variable, the shape and the
size of the vectorial defect is independent of $\gamma$.
Furthermore, we can use the arguments given by Hagan for the
scalar CGL equation \cite{hagan} to determine the wave numbers
$k_+$ and $k_-$. According to this the spiral wavenumber depends
only on the parameters $\alpha$ and $\beta$, therefore $k_+ = k_-
= k_H$, where $k_H(\alpha,\beta)$ is the wavenumber of a spiral in
the scalar CGL. Far from the defect core, as $r \rightarrow
\infty$, $\psi_+(r)=\psi_-(r)=
\psi(r) \rightarrow k_H r$.  The rotation
frequencies are $\omega_+ =\omega_- = \omega$, where $\omega$ is
given by Eq. (\ref{omega}) using $k = k_H$. It turns out
\cite{hagan} that the dependence of $k_H$ on $\alpha$ and $\beta$
is always through the combination
\begin{equation}
\kappa=\left|{\alpha -\beta \over 1+\alpha\beta}\right|\ ,
\label{kappa}
\end{equation}
a parameter that will be important in the following. If $\kappa
=0$, $k_H=0$, so that when $\alpha=\beta$ (the relaxational case)
a homogeneous state is found instead of a wave far from the vector
defect core.

For an argument vectorial defect, $n_+ = n_- = n = \pm 1$, and we
have,
\begin{eqnarray}
A_x & = & \sqrt{2} R(r) \cos(\Delta \phi_0/2)  e^{i[n \theta +
\psi(r) - \omega t + (\phi_{0+} + \phi_{0-})/2]} \nonumber \\
A_y & = & \sqrt{2} R(r) \sin(\Delta \phi_0/2) e^{i[n \theta +
\psi(r) - \omega t + (\phi_{0+} + \phi_{0-})/2]} \ \ ,
    \label{axyarg}
\end{eqnarray}
where $\Delta \phi_0 = \phi_{0+} - \phi_{0-}$ is the difference of
initial phases of $A_+$ and $A_-$. If ${\bf A}$ represents an
optical field, its polarization is linear, since the cartesian
components $A_x$ and $A_y$ oscillate in phase. The angle of
polarization is the constant $\xi = \Delta
\phi_0/2$.

For a director vectorial defect, $n_+ = -n_- = n = \pm 1$, we
have,
\begin{eqnarray}
A_x & = & \sqrt{2} R(r) \cos(n\theta + \Delta \phi_0/2)
e^{i[\psi(r) - \omega t + (\phi_{0+} + \phi_{0-})/2]} \nonumber \\
A_y & = & \sqrt{2} R(r) \sin(n\theta + \Delta \phi_0/2)
e^{i[\psi(r) - \omega t + (\phi_{0+} + \phi_{0-})/2]}.
    \label{axydir}
\end{eqnarray}
The polarization in this case is also linear at each point, but
the angle of polarization, given by $\xi = n\theta +
\Delta \phi_0/2$, rotates around the core of the defect (from here
the name of {\sl director} defect).

The defects just presented do indeed appear spontaneously in
numerical simulations of Eq.~(\ref{vcglexy}) in appropriate
parameter ranges. However, the classification by itself does not
indicate much about the dynamics. It turns out that, when present,
the vectorial defects dominate the dynamics, as shown in
Fig.~\ref{fig:g01}. They generate a large exclusion domain from
which any inhomogeneity or scalar defect is expelled away. The
scalar defects accumulate at the domain limits. The modulus of
$A_+$ and $A_-$ remain essentially frozen in time. The phases
$\phi_+$ and $\phi_-$ display corotating spirals for the argument
defects and counterrotating spirals for the director defects, in
agreement with (\ref{fase}) for the appropriate values of the
charges $n_i$. We also show in Fig.~\ref{fig:g01} the global phase
$\phi_g
=
\phi_+ +
\phi_-$ and the relative phase $\phi_r = \phi_+ - \phi_-$.
Argument and director defects are easily distinguished in the plot
of the global phase: around an argument defect a two-armed spiral
is formed (in a closed path around the defect $\phi_g$ changes in
$4\pi$), while a target pattern is seen in the domain of a
director defect ($\phi_g$ does not change). The relative phase
$\phi_r$ does not change in a closed path around an argument
defect while for a director defect it changes in $4\pi$. The
scalar defects are identified as the points where the global or
relative phase changes by $2 \pi$ in a closed path around the
point. The modulus and phase of $A_x$ are also plotted. In the
domains of argument defects, the modulus of $A_x$ is constant and
the phase is a spiral, in agreement with Eq. (\ref{axyarg}). In
the domains of the director defects, $|A_x|$ has a characteristic
radial structure that arises from the term $\cos(n\theta +
\Delta \phi_0/2)$ in Eq. (\ref{axydir}). The phase of
$A_x$ is a target pattern broken by a straight line where there is
a phase jump of $\pi$. This jump is produced by the change of sign
of the cosine in Eq. (\ref{axydir}).

Figure \ref{fig:scalar_def} shows a scalar defect present in the
$A_+$ component. It has been obtained at the same values of
$\alpha$ and $\beta$ as before, but $\gamma=0.4$. At these
parameter values, vectorial defects do not appear (as discussed
later), so that the scalar ones are allowed to generate wave
domains, as the one shown in Fig.~\ref{fig:scalar_def}. At the
scalar defect core, where one of the component vanishes, the
modulus of the other has a local maximum. Thus the object is
circularly polarized and imposes some elliptic polarization to its
neighborhood. The phase of $A_+$ is a spiral wave whose wavenumber
can not be derived from the Hagan approach, since for scalar
defects we have that $|A_+|
\ne |A_-|$ (see Fig. \ref{fig:scalar_back}) and no reduction to a
single scalar CGL is possible. The wavenumber of the spiral wave
is smaller than $k_H$. The phase of the nonvanishing component is
almost constant. More precisely, this phase slowly changes with
radial symmetry as one moves away from the center, so that the
lines of constant phase form a target pattern. In a similar set of
equations that include convective terms and a complex coupling,
scalar defects have also been reported \cite{lega}.  In that case,
the phase of the nonvanishing field component presents a clearer
target pattern. Phase target patterns in the scalar CGL equation
are also found under suitable boundary conditions
\cite{victor1,victor2} or inhomogeneities \cite{ott_inho}. The
amplitudes and phases of the $x$ and $y$ linearly polarized
components of the field around a scalar defect, also shown in
Figs.~\ref{fig:scalar_def} and \ref{fig:scalar_back}, present
spiral waves. Note that only phase waves, and not amplitude spiral
waves, appear in the scalar CGL equation.

Figure \ref{fig:anticor} shows another view of scalar defects at
two values of $\gamma$. The maximum in the modulus of the
nonvanishing component, associated to the presence of the defect,
is seen to grow with the value of $\gamma$, if $\gamma>0$. This is
a consequence of the coupling term in Eq.~(\ref{vcgle}), which
favors anticorrelation if $\gamma>0$ (but positive correlation if
$\gamma<0$). Notice also that the size of the scalar defect-core
increases with $\gamma$.

\section{Defect dynamics for $|\gamma|<1$}
\label{sect:dynamics}

We first discuss all the theoretically possible processes of
creation and annihilation of defects. These processes are
illustrated in Fig. \ref{fig:esquema2} where we plot in the
horizontal axis the topological charge of the $A_-$ component, and
in the vertical axis the one of $A_+$.  Open circles correspond to
scalar defects, black circles to argument defects and gray circles
to director defects, as in Fig.~\ref{fig:esquema}. The vectorial
sum of arrows in the scheme implements the rule of topological
charge conservation. The possible processes are: i) creation of a
vector defect by the coalescence of two scalar defects
(illustrated in Fig. \ref{fig:esquema2}(a) for an argument
defect). This process is described in subsection
\ref{sect:creation}. ii) Splitting of a vector defect in two
scalar defects, which is in fact the inverse process, (see
subsection \ref{sect:core}). iii) Annihilation of a vector defect
by collision with a scalar one. Fig.~ \ref{fig:esquema2}(b)
represents the annihilation of an argument defect (the black dot)
by collision with a scalar defect in the $A_+$ component (the open
dot in the vertical axis), giving as a result a scalar defect in
the $A_-$ component (the open dot in the horizontal).
Fig.~\ref{fig:esquema2}(c) shows a similar annihilation process
for a director defect (subsection \ref{sect:background}).

Aranson and Pismen performed recently \cite{aranson-pismen} an
analytical study of the interactions between scalar defects at
different field components, aiming at characterizing the stability
properties of vector defects: if two scalar defects in different
components attract each other, a vector defect will be formed in
absence of other processes, whereas vector defects will be
unstable if their components repel. A first result is that the
interaction is long ranged, in contrast to the character of the
interaction between defects in the same field. The approach in
\cite{aranson-pismen} is perturbative for small $|\gamma|$. The
results depend on the value of $\kappa$, the parameter fixing the
asymptotic value of the wavenumber. According to
\cite{aranson-pismen} there is a critical value $\kappa_c \simeq
0.52$ such that if $\kappa >
\kappa_c$ two scalar defects in different components attract each
other for $\gamma>0$ \cite{vectornote}.  For $\gamma<0$, two
scalar defects in different components attract each other if
$\kappa <
\kappa_c$. We find however numerically that for $\gamma < 0$ the
vector defects have a short life time. The reason is that these
defects generate a spiral with a small wavenumber $k_H$ so that
the group velocity at which perturbations are expelled away $v_g =
2(\alpha -\beta)k_H$ (see Appendix) is also small. It is then very
likely that scalar defects in the neighborhood of the vector one
(excluded from the analysis in \cite{aranson-pismen}) approach its
vector core and annihilate one of its components.
In the following we focus our analysis of the dynamical properties
of vector defects to the case $\gamma>0$. Creation, annihilation
and splitting of vector defects are described in the following
three subsections. The relaxational case ($\alpha=\beta$) is
discussed in subsection \ref{sect:relaxational}.

\subsection{Creation of vector defects for $\gamma>0$}
\label{sect:creation}

If $\gamma = 0$ both components of the field behave independently
and vector defects are thus not formed. For $\gamma \rightarrow 0$ the
density of vector defects goes to zero for any value of
$\alpha$ and $\beta$. But the behaviour for increasing $\gamma$
depends on the values of $\alpha$ and $\beta$. It is useful to
have in mind the different stability regions in $\alpha$-$\beta$
parameter space for the spirals in the scalar CGL equation, as
described for example in \cite{aranson}. Spirals simply can be
absent (for $\kappa=0$), or rather be stable, convectively
unstable (the spirals remain in place and look stable because
perturbations are effectively ejected away thanks to the group
velocity on the spiral wave) or absolutely unstable. When the
spirals are absolutely unstable defect turbulence develops.
The extent of these regions is altered here by the coupling between
the fields, given by the parameter $\gamma$.

If $\kappa \neq 0$, for a coupling $0 < \gamma < \gamma_d$ (where
$\gamma_d$ depends on $\alpha$ and $\beta$), vectorial defects are
formed at short times starting from a random initial condition.
Initially there is a high density of scalar defects.  The density
decreases as defects of opposite charges collide and annihilate in
pairs in each component of the field. At this point, two scalar
defects, one in $A_+$ and the other in $A_-$, may get close to
each other, join and form a bound structure.  This later stage is
reached later as $\gamma$ becomes smaller.  Figure
\ref{fig:genesis} shows the creation of a vectorial defect by
displaying the sum of the field amplitudes, $|A_+|^2 + |A_-|^2$
($\alpha = 0.2$, $\beta = 2$, so that $\kappa=1.29>\kappa_c
\approx 0.52$, and $\gamma=0.1$). For these values of $\alpha$ and
$\beta$ the scalar CGL (\ref{cgle}) is in the region where the
wavenumber $k_H$ of the spirals is convectively unstable
\cite{aranson}. In the temporal sequence it can be seen that two
dark dots, corresponding to the two scalar defects, one in each
component of the field, get together and finally form the vector
structure (the black dot). Immediately after the coalescence of
both defects, the group velocity around the newly formed vector
defect produces an exclusion region that precludes the
approximation of other scalar defects. The region grows until the
system arrives to a configuration similar to the one displayed in
Fig.~\ref{fig:g01}. The system has reached a frozen structure of
domains separated by shock waves. The configuration in
Fig.~\ref{fig:g01} is qualitatively similar to the ones
encountered for the scalar CGL. Since very slow relaxation an
other similarities with structural glasses have been pointed out
in that case, the terms {\sl glassy state} or {\sl glass phase}
are applied to it. We will use also these terms for the vectorial
case (which is also seen to evolve in a very slow time scale). The
peculiarity here is that, whereas in the scalar CGL the difference
between the central defect and the defects at its domain border
seems to arise from a spontaneous symmetry breaking, here all
vector defects always expel from its domain the scalar ones, so
that vector defects, if present, are always found at the center of
wave domains.

The glassy configurations occur for relatively small $\gamma$. As
$\gamma$ is increased vectorial defects become unstable, leading
to the melting of the glass phase. We have identified two
mechanisms by which this process occurs, to which the next two
subsections are devoted: (i) domain instability and (ii) core
instability. In the case of domain instability the domain around a
vector defect becomes unstable and develops inhomogeneities. These
irregularities lead to paths that permit the approach of a scalar
defect that finally collides with the core and annihilates one of
the components of the vector defect. In the core instability case
the vector defect simply splits in two scalar defects. This
process is described in Ref.~ \cite{pismen} for the real
coefficient case, where a greater symmetry between director and
argument defects seems to be present and both kind of defects
become unstable for the same value of $\gamma$.  As we will see in
Sect. \ref{sect:core}, this is not the case for complex
coefficients.

Finally, in the region of parameter space $\alpha$-$\beta$ where
spirals are absolutely unstable \cite{aranson} for the scalar CGL
equation, no vectorial defects are formed. In this case we have
what is called in the scalar case defect turbulence. Increasing
$\gamma$ does not change the qualitative behavior of each
component, although anticorrelation between the modulus of both
components increases.

\subsection{Domain instability}
\label{sect:background}

By `domain instability' we denote the situation in which the wave
domain around a vector defect becomes unstable, begins to
fluctuate, and develops inhomogeneities. Shortly afterwards one of
the two charges forming a vector defect is annihilated by an
external scalar defect. As a result a free scalar defect is left
in the other component of the field. The region of parameter space
$\alpha$-$\beta$ where this process is observed corresponds
approximately to the region where for the scalar CGL equation the
phase spirals are convectively unstable \cite{aranson}.
As $\gamma$ is increased from zero the stability of
the spirals is modified: At a given value of $\gamma$, the group
velocity is not strong enough to overwhelm the growth of the
perturbations, the spirals becoming absolutely unstable. At this
point the domains around the vectorial defects are uneffective as
exclusion zones, so that scalar defects previously confined to the
domain border can approach the vectorial defect core. Although
this picture is valid for both kinds of vectorial defects,
director defects survive for larger $\gamma$ than argument ones.
For the parameter values of Fig.~\ref{fig:g01}, argument defects
become unstable at $\gamma
\simeq 0.3$ (see Fig. \ref{fig:aniq_argum}), while director defects
remain stable up to $\gamma \simeq 0.35$ (see Fig.
\ref{fig:aniq_dir}). For larger $\gamma$ only scalar defects are
found numerically.

The different stability range of argument and director defects can
be understood through a linear stability analysis of the vector
spirals focusing in its plane-wave structure far from the core.
This analysis is performed in the Appendix. The main point to be
recognized is the difference between the phase structure between
argument and director vector defects. In Figs. \ref{fig:lin_fase}
(a) and (b) we show constant phase curves of both components for
an argument defect and a director defect. The wave vectors of the
components $A_+$ and $A_-$ in a point ${\bf x}$, ${\bf k_+}({\bf
x})$ and ${\bf k_-}({\bf x})$ are perpendicular to the
constant-phase curves. As discussed before, $k_+ = k_- = k_H$,
where $k_H$ is the wave number of a scalar spiral. For the
argument defects the difference of wave vectors ${\bf k_R} = {\bf
k_+} - {\bf k_-}$ vanishes at any point of the plane while for the
director defects ${\bf k_R} \neq 0$. Far from the defect core,
${\bf k_R}$ decreases for the director defect, however it remains
different from 0 within the exclusion island surrounding the
defect.

The linear stability analysis of the wave in the two cases turns
out to be different. It is shown in the Appendix that for argument
defects, a wave with $k_+=k_-=k_H$ becomes unstable if $k_H>K_p$,
where $K_p(\alpha,\beta,\gamma)$ is given in Eq.~(\ref{kp}),
while in the case of director defects this happens if $k_H>K_s$,
where $K_s(\alpha,\beta,\gamma)$ is given in Eq.~(\ref{ks}).

In figures \ref{fig:ab} (a) and (b) we plot stability diagrams in
the $\alpha$-$\beta$ plane, with the line $1 + \alpha \beta = 0$
as a reference. In fig. \ref{fig:ab} (a) we show the curves $K_p =
k_H$ and $K_s = k_H$ for $\gamma = 0.3$. To the right of the
curves, plane waves with $k = k_H$ are unstable. In fig.
\ref{fig:ab} (b) we plot the curve $K_p = k_H$ for different
values of $\gamma$ ($\gamma = 0, 0.3, 0.6, 0.9, 0.99$). We can see
that the curve moves left and upwards as $\gamma$ is increased;
the same happens with the curve $K_s = k_H$.  Let us consider a
point $(\alpha_0,\beta_0)$ such that plane waves with $k=k_H$ are
stable for $\gamma=0$.  As $\gamma$ is increased the curve $K_p =
k_H$ crosses the point turning the vectorial waves with ${\bf k_R}
= 0$ unstable, this destabilizes the domains of the argument
defects. If $\gamma$ is further increased, then the curve  $K_s =
k_H$ crosses the point $(\alpha_0,\beta_0)$ and waves with ${\bf
k_R}
\neq 0$ become unstable, destabilizing the domains of the director
defects. It should be said that the kind of stability analysis
performed in the Appendix considers extended perturbations. Since
spiral waves have a group velocity, it may happens that these
instabilities are of convective type, that is, localized
perturbations do not destroy the spiral wave but they are advected
away from the defect core of the spiral with the group velocity
\cite{aranson}. Thus, we can still find stably-looking spirals
although the corresponding plane waves far from the core are
unstable. The calculation of the limit of {\sl absolute
instability} to localized perturbations is quite involved, but the
result for the convective instability suggests that corotating
spirals become absolutely unstable before counterrotating ones. We
believe that the analysis presented here is an indication of the
kind of instabilities that can affect the vectorial spirals and
the order in which they appear as the coupling parameter $\gamma$
is increased.


\subsection{Core instability}
\label{sect:core}

A core instability of a vector defect occurs when it splits in two
scalar defects. This process is roughly present in the region of
parameter space $\alpha$-$\beta$ where the spirals are stable in
the scalar case. The splitting of a director defect is shown in
Fig.~\ref{fig:split_dir} for $\kappa = 0.54$ and $\gamma=0.95$.
The size of the vectorial defect-core is much narrower than the
size of the core of the two scalar defects that remain at the end
of the process. This is a consequence of two facts mentioned
before, namely that the size of the vectorial defects does not
depend on $\gamma$, whereas the size of scalar defects increases.
Also in this case argument defects become unstable for smaller
$\gamma$ than director defects. For example, for the parameter
values of Fig.~\ref{fig:split_dir} argument defects already split
for $\gamma = 0.75$.

After the splitting, both scalar defects may form a bound pair
that resembles a rotating molecule \cite{aranson-pismen}. In Fig.
\ref{fig:rota} a temporal sequence is plotted that shows a pair of
scalar defects that rotate one around the other.  For the
particular values of the figure, however, the distance between the
scalar defects increases and the angular velocity decreases, thus
the molecule finally unbinds. Stable molecules, however, are found
for other parameter values as reported in \cite{aranson-pismen}.

\subsection{The relaxational case: $\alpha=\beta$}
\label{sect:relaxational}

Qualitative features distinguish the case $\alpha=\beta$ (
$\kappa=0$) from the other cases: First, no spiral wave is formed
around vector defects (since, as stated before, $k_H=0$ if
$\kappa=0$). Second, the group velocity with which linear
perturbations are expelled away from the neighborhood of a vector
defect is zero (the group velocity is $v_g=2(\alpha-\beta)k_H$, as
shown in the Appendix). Thus, perturbations and other defects are
rather free to come arbitrarily close to vector defects if
$\kappa=0$ and we expect them to be frequently annihilated by
processes such as the ones shown in Fig.~\ref{fig:esquema2}(b) or
(c).

Starting from random initial conditions we do not find numerically
the formation of vectorial defects in Eq. (\ref{vcgle}) for
$\alpha=\beta$ and $\gamma>0$. In Fig. \ref{fig:real}, the
quantities $|A_+|^2$ and $|A_-|^2$ are plotted for $\alpha=\beta=0$
and $\gamma = 0.1$. It can be seen that there is no superposition
of two defects in the same point. Starting with an initial
condition with a slightly perturbed vector defect, it spontaneously
splits into two scalar defects for any positive $\gamma$, as can be
seen in Fig. \ref{fig:split} for $\gamma = 0.2$. This behavior is
in agreement with the predictions of \cite{aranson-pismen}.

More in general, approaching the line $\alpha=\beta$, we observe
numerically that director and argument defects of initial
configurations such as the one in Fig.~\ref{fig:g01} split
spontaneously, even for very small values of $\gamma$, again in
agreement with \cite{aranson-pismen}.

\section{The gas phase}
\label{sect:gas}

For $\gamma$ high enough, the vectorial defects always disappear
following one of the two mechanisms described above. The system
then presents a faster disordered dynamics compared with the {\it
glassy} phase with vector defects.  It is a kind of {\em gas-like}
phase, dominated by the scalar defects, which are conserved in
number during very long times. The scalar defects move faster than
in the glassy phase, in a way reminiscent of atoms in a gas, from which
we borrow the name. Furthermore, the
spiral wavelength around scalar defects increases with $\gamma$,
so that well-developed spirals do not fit in the domains for
$\gamma$ close to one. Domains are thus less effective as
exclusion zones and defects even more mobile.

To distinguish the presence of these two different phases we can
use the joint probability distribution of the modulus of the field
components: $p(|A_+|,|A_-|)$.  Figure \ref{fig:joint_prob} shows a
grey scale plot of $p(|A_+|,|A_-|)$ for $\gamma = 0.1$ (glassy
phase) and $\gamma = 0.95$ (gas phase); in both cases $\alpha=0.2$
and $\beta=2$.  The plot is obtained taking the values of
$(|A_+|$,$|A_-|)$ at different space-time points.  For $\gamma =
0.1$ there is a broad maximum around $|A_+| \simeq |A_-|
\simeq 0.85$ with deviations from these values being rather uncorrelated,
except for the points lying in the line $|A_+|=|A_-|$. This line
shows that the absolute values of both components take
simultaneously the same value between 0 and 1, situation
identifying the core of vector defects.  For $\gamma = 0.95$ the
line $|A_+|=|A_-|$ has disappeared due to the absence of vector
defects.  Instead, the probability distribution approaches the
curve given by $|A_+|^2 + |A_-|^2 = 1$, which indicates
anticorrelation between $|A_+|$ and $|A_-|$. This anticorrelation
is the fingerprint of the dominance of the scalar defects (See
Figs.~\ref{fig:scalar_back} and \ref{fig:anticor}). Similar
qualitative behaviour is observed for other values of $\alpha$ and
$\beta$ (given that $\kappa > \kappa_c$) as $\gamma$ is increased.

The behavior in the glass and in the gas phase can be interpreted
in terms of synchronization and generalized synchronization,
respectively, of spatiotemporally chaotic configurations of the
two field components \cite{amengual,hernandez}. In this context,
another interesting quantity that gives information about the
transition from the glassy to the gas phase is the mutual
information between field components, that can be computed from
the individual and joint probability densities. In general, for
two random discrete variables $X$ and $Y$ the mutual information
$$I(X,Y) = -\sum_{x,y} p(x,y) \ln
\frac{p(x)p(y)}{p(x,y)}$$
gives a measure of the statistical dependence between both
variables, the mutual information being 0 if and only if $X$ and
$Y$ are independent \cite{mc}. Numerical results presented in
\cite{hernandez} show that there is a minimum of $I(|A_+|,|A_-|)$
($\gamma \simeq 0.3$ at $\alpha = 0.2$ and $\beta = 2$), where the
variables have been adequately discretized, and that this value of
$\gamma$ can be interpreted as the transition point from the
glassy to the gas phase. See \cite{hernandez} and \cite{hoyuelos}
for more details about the vortex unbinding transition between the
glassy and the gas-like phases.

\section{Dynamics for $\gamma>1$}
\label{sect:gamma1}

For  $\gamma > 1$, as discussed in Sec. \ref{sect:eqs}, linearly
polarized states become unstable with respect to circularly
polarized states. In this case the system segregates in two
phases, one right circularly polarized and the other left
circularly polarized. Figure \ref{fig:g11} is an example of a
configuration that started from a random initial condition
($\gamma=1.1$, $\alpha=0.2$, $\beta=2$). Phase separation is
clearly observed. As the system evolves, the length of the
interface walls is reduced, in a domain-coarsening process.
Despite the tendency of the concave domains to shrink, some
defects remain at long times. Vector point defects are no longer
topologically allowed since the stable uniform states are such
that the vanishing of one of the two polarizations in large areas,
not just at points, is favored. However we still have scalar point
defects that are now lying on the domains of one circular
polarization, and are points at which the corresponding field
amplitude goes to zero. Only one topological charge is defined,
the associated to the component which vanishes just at one point.
They are of the type described in \cite{pismenPRL,pismen} as
having a `repolarized core' structure: In domains filled with one
of the polarizations, say $+$, a defect is a place at which
$|A_+|$ goes to zero, thus producing a phase singularity (seen in
Fig.~\ref{fig:g11}) associated to a topological charge $n_+=\pm
1$; in response to this behavior the component which does not
carry the topological charge takes nonzero values in the
defect-core region. Other structures for the core are in principle
possible, as for example a `punched core' structure in which the
component that does not carry the charge remains vanishing in the
defect-core region. As discussed in \cite{pismenPRL,pismen},
topological arguments imply that only the `repolarized core'
should be observed for $\gamma$ slightly above 1, and this is
indeed what Fig.~\ref{fig:g11} shows. The differences in the sizes
of the defect cores seen in Fig.~\ref{fig:g11} disappear at longer
times.

We stress that the topological defects in Fig.~\ref{fig:g11} are
different from the localized structures found in other systems
with coexistence of two equivalent homogeneous states, such as the
Swift-Hohenberg equation in adequate parameter ranges or the
parametrically forced CGL equation. In these systems the
oscillating tails of the fronts connecting the two equivalent
homogeneous solutions are responsible for maintaining the
existence of localized structures against the tendency of the
domains to shrink \cite{gomila}. These localized structures are
not topological defects. Here, on the contrary, it is the
topological character of the objects what prevents them to
disappear after shrinking to a point: defects can only disappear
by combination with others of opposite charge.

\section{Conclusions}
\label{sect:conclusions}

We have numerically explored the conditions for the existence and
stability of the different kinds of defects in the twodimensional
VCGL equation: vector defects of argument or director type, and
scalar defects. Dynamic phenomena associated with their creation
and annihilation processes have been described. In particular, we
have identified two mechanisms that lead to the destruction of the
vector defects as the coupling parameter between field components
is increased: domain instability and core instability. One or the
other takes place depending on parameters $\alpha$ and $\beta$.
For the domain instabilities of vector defects we have obtained
analytic understanding in terms of the convective instability of
the emitted spiral wave. For the core instability at a relatively
large threshold value of $\gamma$ there is no available theory so
far.

Our numerical results have been compared with the useful
analytical studies by Aranson and Pismen \cite{aranson-pismen}.
They studied core instabilities as a function of $\alpha$ and
$\beta$ for small $\gamma$. Their results explain our result that
in the relaxational limit ($\alpha=\beta$) there are no vectorial
defects for $\gamma>0$. They also explain the existence of vector
defects far from the relaxational limit and for small $\gamma$.
Our main interest, however, was on what happens far from the
relaxational limit as function of $\gamma$. In this regime we have
found threshold values of $\gamma$ for the domain and core
instabilities mentioned before. These instabilities are not
accessible to the perturbative theoretical discussion of Aranson
and Pismen and there is need of further analytical studies.

Finally, the dominance of different types of defects at different
parameter values leads to the identification of different
dynamical regimes, such as the glassy and the gas phase.

\section*{Acknowledgments}

Financial support from MCyT (Spain), project CONOCE BMF2000-1108,
is acknowledged. M. H. acknowledges financial support of
Fundaci\'{o}n Antorchas, Argentina, and of the National Council
for Scientific and Technical Research (CONICET) from Argentina.

\section*{Appendix}

We analyze in this Appendix the stability of the wave emitted by
vector defects. Far from the defect core we can approximate the
spiral waves by plane waves. The stability of plane waves with wave
number $k_H$ will give us information about the stability of the
spirals. The stability analysis presented here is an extension to
two dimensions of the analysis made in one dimension in
\cite{sanmiguel} and it is valid for $|\gamma|<1$. The effect of a
perturbation on the plane wave solution (\ref{plana1}) can be
written as
\begin{equation}
A_\pm = (Q + a_\pm) e^{-i({\bf k_\pm} \cdot {\bf r} - \omega_\pm t
+
\phi_{0\pm})}
    \label{plane_wave}
\end{equation}
where $|{\bf k_+}|=|{\bf k_-}| =k_H$ for a vector defect. $Q$ and
$\omega$ are given by Eq. (\ref{omega}), and $a_\pm$ are the
complex perturbations. Using the variables
\begin{equation}
 \left \{ \begin{matrix}
{s\cr s_I\cr r\cr r_I}\end{matrix}
\right \} =
\left \{ \begin{matrix}
{\Re(a_+ + a_-)\cr
\Im(a_+ + a_-)\cr
\Re(a_+ - a_-)\cr
\Im(a_+ - a_-)}\end{matrix}
\right \},
\end{equation}
where $\Re$ and $\Im$ mean real and imaginary part, respectively,
we obtain linearized equations for $s$, $s_I$, $r$ and $r_I$.
Instabilities appear first in the phase modes. They are identified
as the linear combinations of the variables that in the
space-homogeneous case lead to zero eigenvalues. They are $\theta
= -\beta s + s_I$ and $\psi = -\beta r + r_I$.  Performing the
change of variables $(s,r,s_I,r_I)
\rightarrow (s,r,\theta,\psi)$, and eliminating adiabatically the
rapidly decaying amplitude variables $s$ and $r$ in terms of the
phases, we get the following phase equations:
\begin{eqnarray}
\partial_t \theta &=& (\alpha - \beta) {\bf k_S} \cdot \nabla \theta + (1 + \alpha
\beta) \nabla^2 \theta - \frac{(1 + \beta^2)}{2(1 - k^2)} \left [ ({\bf k_S}
\cdot \nabla)^2 \theta + \frac{1 + \gamma}{1 - \gamma} ({\bf k_R} \cdot \nabla)^2
\theta \right ] \nonumber \\
&& + (\alpha - \beta) {\bf k_R} \cdot \nabla \psi - \frac{(1 +
\beta^2)}{(1 - k^2)(1 - \gamma)} ({\bf k_S} \cdot \nabla)({\bf k_R}
\cdot \nabla) \psi, \nonumber\\
\partial_t \psi &=& (\alpha - \beta) {\bf k_S}\cdot \nabla \psi + (1 + \alpha
\beta) \nabla^2 \psi - \frac{(1 + \beta^2)(1+\gamma)}{2(1 - k^2)(1-\gamma)}
\left [ ({\bf k_S} \cdot
\nabla)^2 \psi + \frac{1 - \gamma}{1 + \gamma} ({\bf k_R} \cdot \nabla)^2 \psi
\right ] \nonumber \\
&& + (\alpha - \beta) {\bf k_R} \cdot \nabla \theta - \frac{(1 +
\beta^2)}{(1 - k^2)(1 - \gamma)} ({\bf k_S} \cdot \nabla)({\bf k_R} \cdot \nabla)
\theta,
    \label{lin_eqs}
\end{eqnarray}
where ${\bf k_S} = {\bf k_+} + {\bf k_-}$ and ${\bf k_R} = {\bf
k_+} - {\bf k_-}$.

As explained in Sect.~\ref{sect:background}, ${\bf k_R}=0$ in the
domain surrounding argument defects, but ${\bf k_R} \neq 0$,
although small, for the domain around director defects. In both
cases one can approximate $|{\bf k_S}| \approx 2 k_H$. For small
${\bf k_R}$, the dominant first-order derivative terms in
(\ref{lin_eqs}) indicate that small perturbations on the waves
travel at a group velocity $v_g=2(\alpha-\beta)k_H$.

We consider perturbations of small wave vector ${\bf q}$. The
eigenvalues of the linear system (\ref{lin_eqs}) are,
\begin{equation}
\frac{\lambda_\pm}{q^2} = -1 - \alpha \beta + \frac{(1 +
\beta^2)}{2Q^2 (1-\gamma^2)}\left[ ({\bf k_R} \cdot \hat{\bf q})^2 +
({\bf k_S} \cdot \hat{\bf q})^2 \right]  \pm \frac{\sqrt{S}}{2q}
    \label{autoval}
\end{equation}
where $q=|{\bf q}|$, $\hat{\bf q}={\bf q}/q$ is the direction of
the perturbation vector, and $S = a q^2 + i b q - c$, with,
\begin{eqnarray}
a &=& \frac{(1 + \beta^2)^2}{Q^4 (1-\gamma^2)^2}\left\{\gamma^2
\left[({\bf k_R} \cdot \hat{\bf q})^2 - ({\bf k_S} \cdot \hat{\bf q})^2\right]^2 +
4({\bf k_R} \cdot \hat{\bf q})^2({\bf k_S} \cdot \hat{\bf q})^2
\right\} \nonumber \\
b &=& 8(\alpha-\beta) \frac{(1 + \beta^2)}{Q^2 (1-\gamma^2)}({\bf k_R}
\cdot \hat{\bf q})^2 {\bf k_S} \cdot \hat{\bf q} \nonumber \\
c &=& 4(\alpha-\beta)^2({\bf k_R} \cdot \hat{\bf q})^2 \ \ .
\end{eqnarray}
If ${\bf k_R} \neq 0$, as is the case for director defects, we can
approximate $\sqrt{S} = i \sqrt{c}(1 - \frac{i b}{2c} q) + O(q^2)$.
The real part of the two eigenvalues are,
\begin{equation}
\frac{\Re (\lambda_\pm)}{q^2} = -1 - \alpha \beta + \frac{2(1 +
\beta^2)}{Q^2 (1-\gamma^2)} k^2 \cos^2 \chi_\pm + O(q),
    \label{autoval1}
\end{equation}
where $\chi_\pm$ are the angles between the perturbation wave
vector ${\bf q}$ and the wave vectors ${\bf k_+}$ and ${\bf k_-}$.
Since $|{\bf k_R}|$ is finite the instability may arise for
perturbation wavenumber $q$ {\sl smaller} than $|{\bf k_R}|$. In
this case, considering the most dangerous longitudinal
perturbations, i.e. when $\chi_+ = 0$ or $\chi_- = 0$, we obtain a
critical wave number $K_s$ such that if $k > K_s$ the plane waves
become Eckhaus unstable.  Taking $\Re (\lambda_\pm)=0$ in
(\ref{autoval1}) we get,
\begin{equation}
K_s^2 = \frac{(1 + \alpha \beta)(1 - \gamma)}{(1 + \alpha \beta)(1
- \gamma) + 2(1 + \beta^2)}.
    \label{ks}
\end{equation}
The equations for $\partial_t \theta$ and $\partial_t \psi$
(\ref{lin_eqs}) are coupled, and the eigenvector corresponding to
the unstable eigenvalue mixes both variables.  This means that the
sum and the difference of the field perturbations $a_+$ and $a_-$,
become unstable simultaneously in this case.

If ${\bf k_R} = 0$, as appropriate around argument defects, we
have that ${\bf k_+} = {\bf k_-}$, and $\chi_+ =
\chi_- = \chi$, and
\begin{equation}
S=\left(  { (1+\beta) \gamma k q \cos\chi   \over   Q^2
(1-\gamma^2) }
\right)^2  \ \ .
\end{equation}
We obtain for the real parts of the eigenvalues,
\begin{equation}
\frac{\Re (\lambda_\pm)}{q^2} = -1 - \alpha \beta + \frac{2(1 + \beta^2)(1 \pm
\gamma)}{Q^2 (1-\gamma^2)} k^2 \cos^2 \chi + O(q).
    \label{autoval2}
\end{equation}
 We consider again longitudinal perturbations
corresponding to $\chi = 0$. Taking the marginal stability
condition in (\ref{autoval2}) for the most dangerous eigenvalue,
$\Re (\lambda_+)=0$, we see that in this case plane waves become
unstable for $k > K_p$, where $K_p$ is given by,
\begin{equation}
K_p^2 = \frac{(1 + \alpha \beta)(1 - \gamma)}{(1 + \alpha \beta)(1
- \gamma) + 2(1 + \beta^2)(1 + |\gamma|)}.
    \label{kp}
\end{equation}
For ${\bf k_R} = 0$, the equations (\ref{lin_eqs}) for $\partial_t
\theta$ and $\partial_t \psi$  are decoupled.  The most
restrictive eigenvalue, $\lambda_+$, is associated with the
variable $\psi$, which is related to the difference of the field
perturbations.  When $\psi$ becomes unstable, the difference $a_+
- a_-$ grows; in optical language this is called a polarization
instability.

\newpage

\begin{center}

\begin{figure}
\psfig{figure=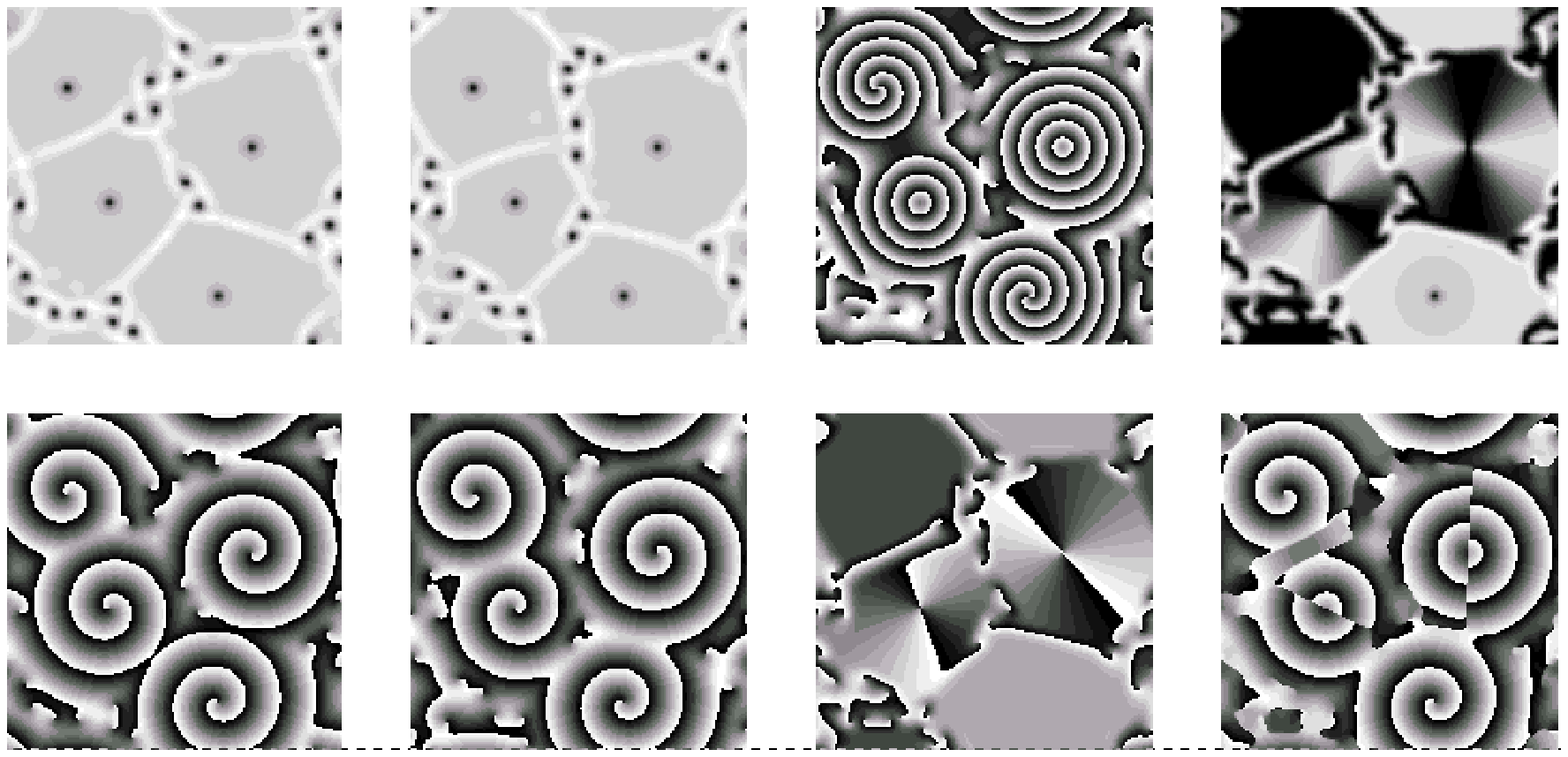,width=14cm} \vspace*{3cm} \caption{Vectorial
defects for $\gamma = 0.1$, $\alpha=0.2$ and $\beta=2$
($\kappa=1.29$). From left to right, top row: $|A_+|^2$,
$|A_-|^2$, global phase $\phi_g$, and $|A_x|^2$. Bottom row:
$\phi_+$, $\phi_-$, relative phase $\phi_r$, and phase of $A_x$.
Here and in the following figures, the values are coded in gray
scale. In the images displaying amplitude values, black means zero
amplitude, and thus the presence of a defect. } \label{fig:g01}
\end{figure}

\begin{figure}
\psfig{figure=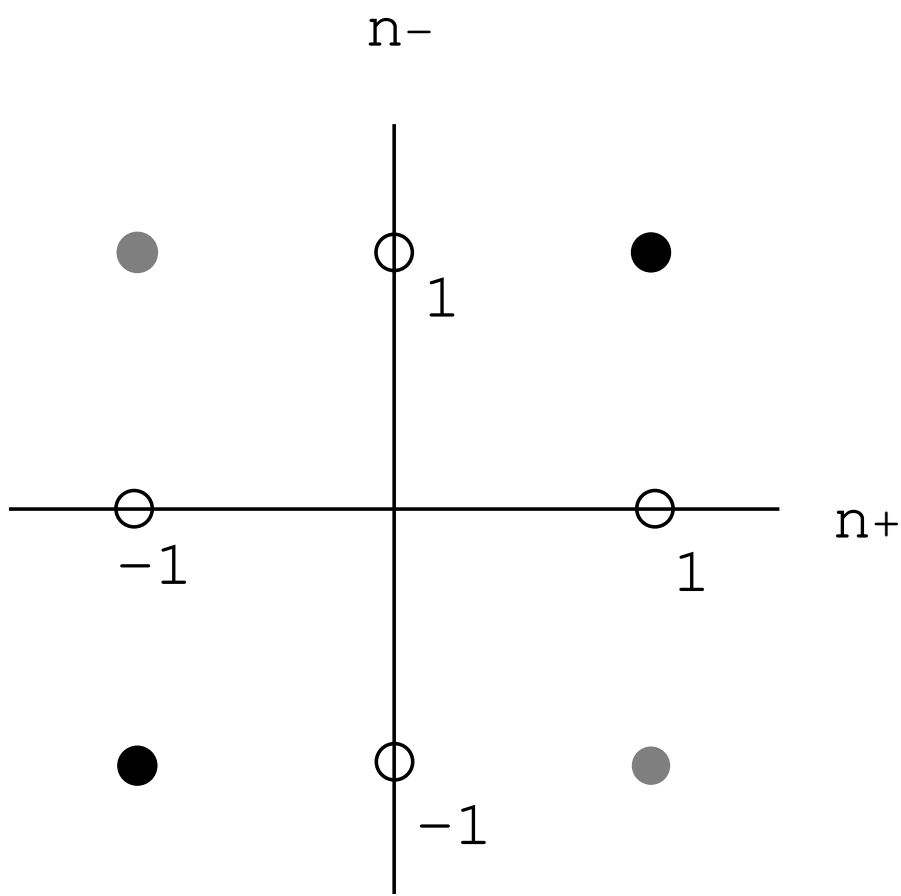,width=12cm} \vspace{1cm} \caption{Scheme
showing the possible defects according to different combination of
the charges $n_+$ and $n_-$.  White dots correspond to scalar
defects, black dots to vectorial argument defects, and gray dots
to vectorial director defects.} \label{fig:esquema}
\end{figure}

\newpage
\begin{figure}
\vspace*{2cm} \psfig{figure=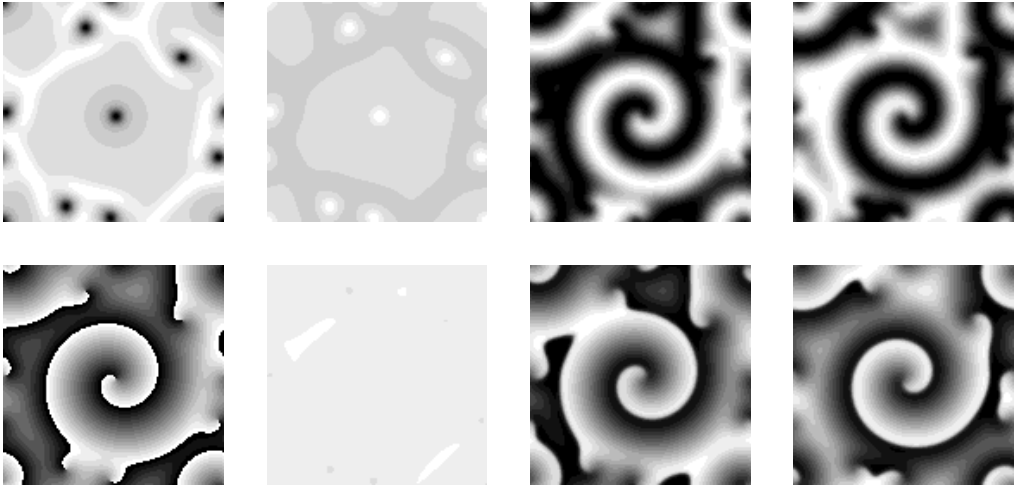,width=14cm}
\vspace*{8cm} \caption{Different views of a scalar defect for
$\gamma = 0.4$, $\alpha=0.2$ and $\beta=2$ ($\kappa=1.29$), present
in the $A_+$ component.  From left to right, top row: $|A_+|^2$,
$|A_-|^2$, $|A_x|^2$ and $|A_y|^2$; bottom row:  $\phi_+$,
$\phi_-$, and phases of $A_x$ and $A_y$,} \label{fig:scalar_def}
\end{figure}

\begin{figure}
\psfig{figure=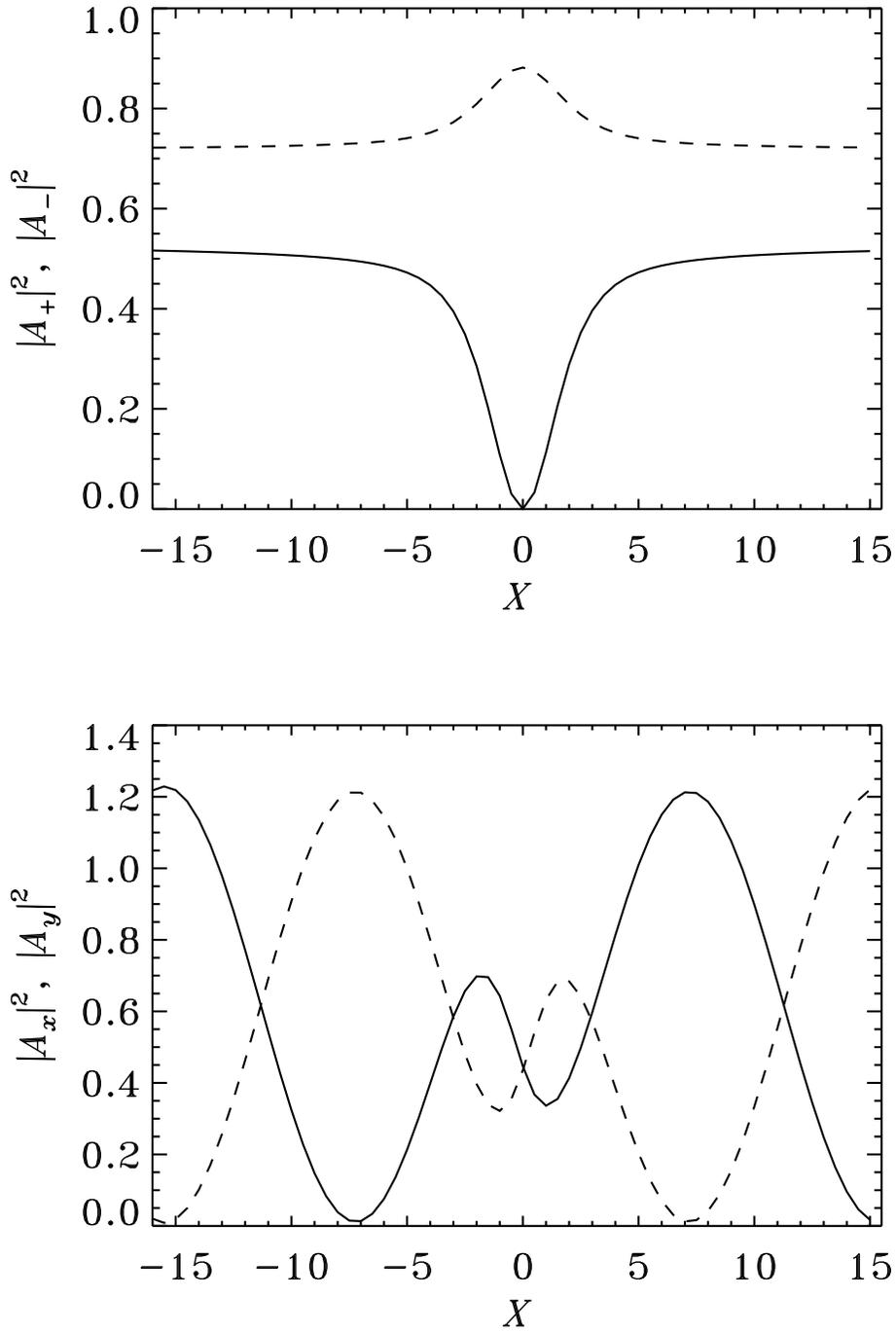,width=14cm} \caption{Cross sections of
the intensities of a scalar defect (same parameters as in Fig.
\ref{fig:scalar_def}).  The top figure shows the defect in the $+$
component (solid line), and associated maximum in $|A_-|^2$
(dashed line). Bottom figure: the spiral waves in the $x$ and $y$
amplitudes are out of phase (compare with Fig.
\ref{fig:scalar_def}).}
\label{fig:scalar_back}
\end{figure}

\begin{figure}
\psfig{figure=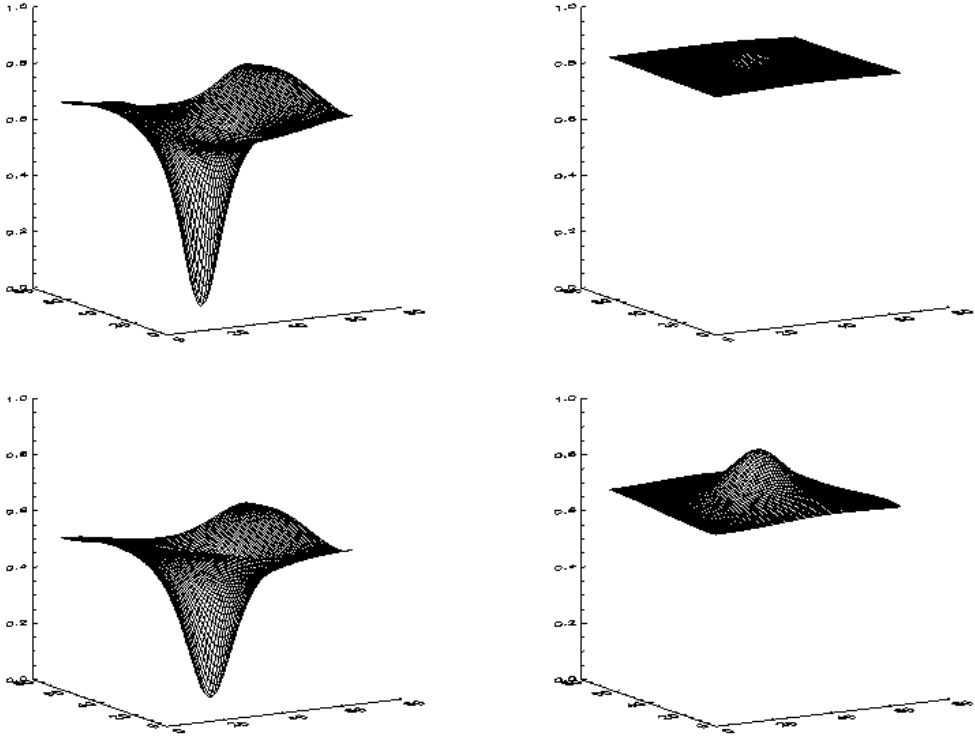,width=14cm} \caption{Scalar defects at different
values of $\gamma$, as a function of the two spatial coordinates.
First row: $|A_+|^2$ (left) and $|A_-|^2$ (right) at $\gamma =
0.1$. Second row: Idem for $\gamma = 0.4$. In both cases
$\kappa=1.29$, and the singular component is $A_+$.}
\label{fig:anticor}
\end{figure}

\begin{figure}
\psfig{figure=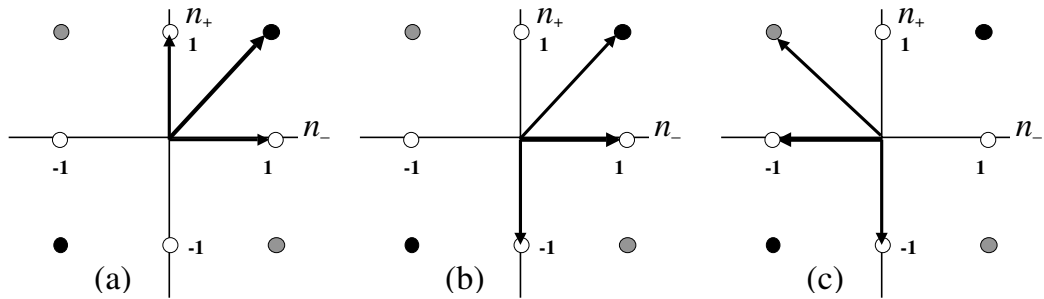,width=14cm}
\caption{Scheme showing some possible creation and destruction
processes of defects in the $n_+$-$n_-$ plane. See text for
details. (a) Creation of an argument defect, (b) annihilation of
an argument defect due to the collision with a scalar defect with
charge $n_-=-1$, (c) annihilation of a director defect.}
\label{fig:esquema2}
\end{figure}

\begin{figure}
\vspace*{10cm} \psfig{figure=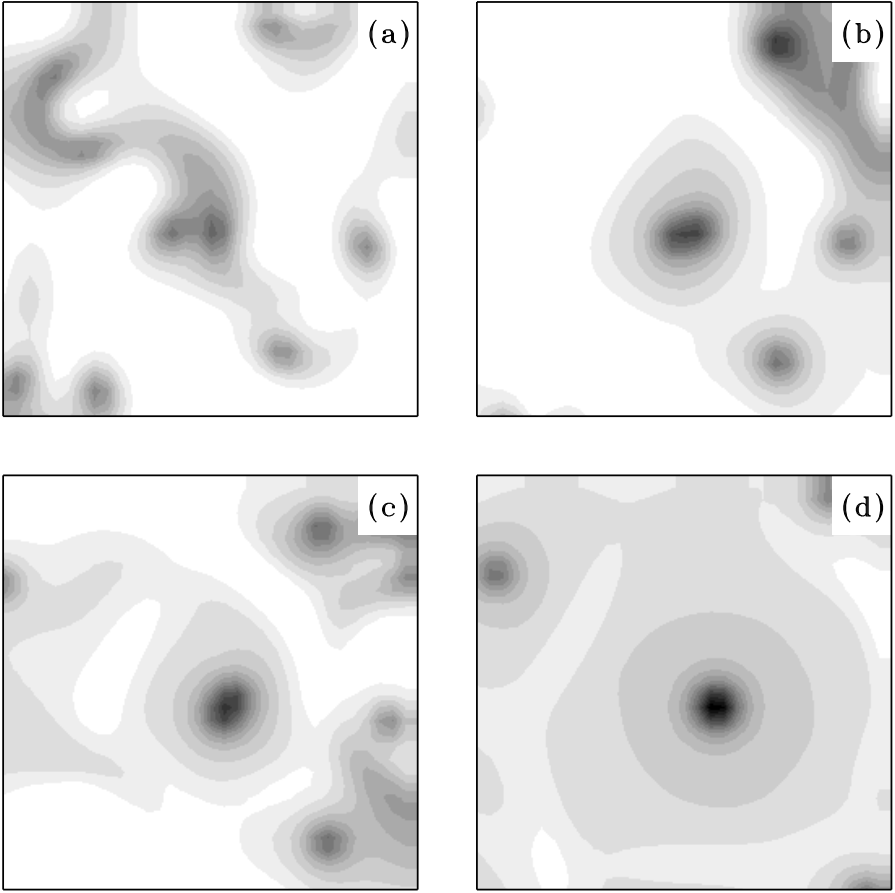,width=14cm}
\vspace*{-3cm}
\caption{Temporal sequence of $|A_+|^2 + |A_-|^2$ showing the
formation of a vectorial defect at short times for $\gamma = 0.1$
and $\kappa=1.29$. (a) $t=15$, (b) $t=30$, (c) $t=45$, and (d)
$t=150$.}
\label{fig:genesis}
\end{figure}

\begin{figure}
\psfig{figure=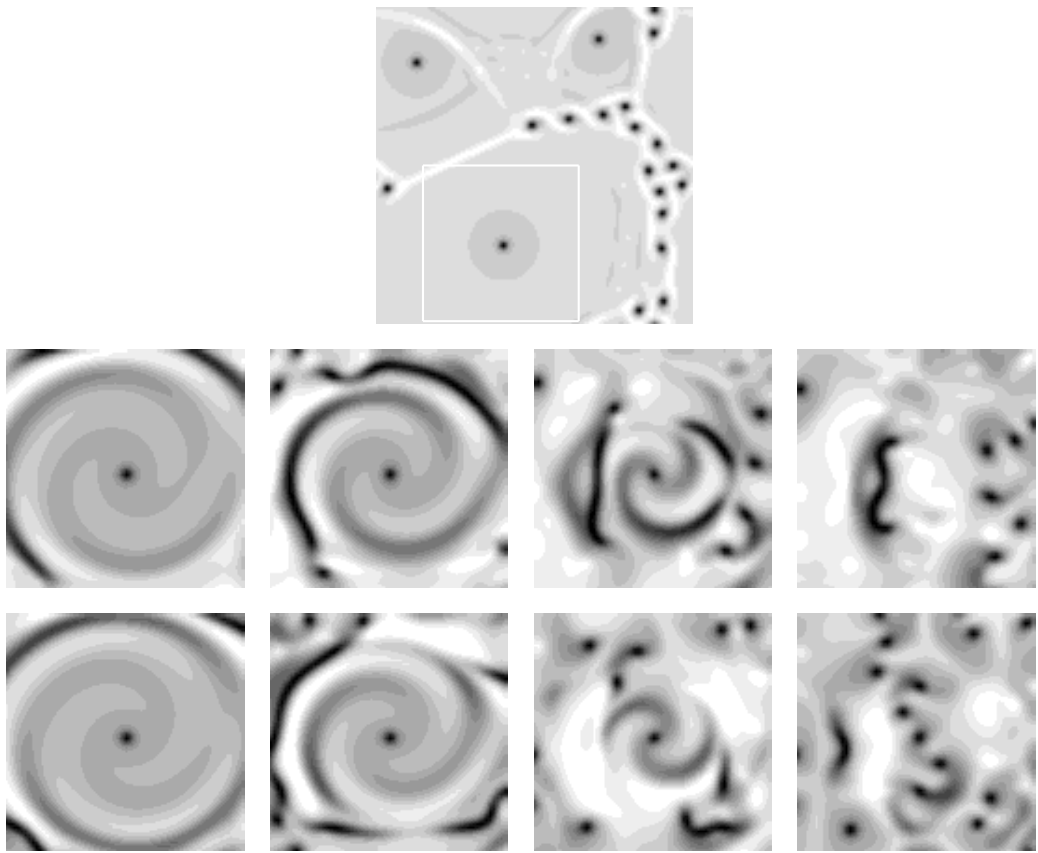,width=14cm} \vspace{1cm}
\caption{Annihilation of an argument defect, for $\gamma = 0.3$ and
$\kappa = 1.29$. The top figure shows the initial condition
generated with $\gamma=0.1$, the box indicates the domain of the
argument defect, and is amplified in the images below. First row:
$|A_+|^2$, second row: $|A_-|^2$. From left to right: $t=t_0+30$,
$t=t_0+60$, $t=t_0+90$, and $t=t_0+120$.  At the final stage
scalar defects approach the defect core annihilating the $n_+$
charge.}
\label{fig:aniq_argum}
\end{figure}

\begin{figure}
\psfig{figure=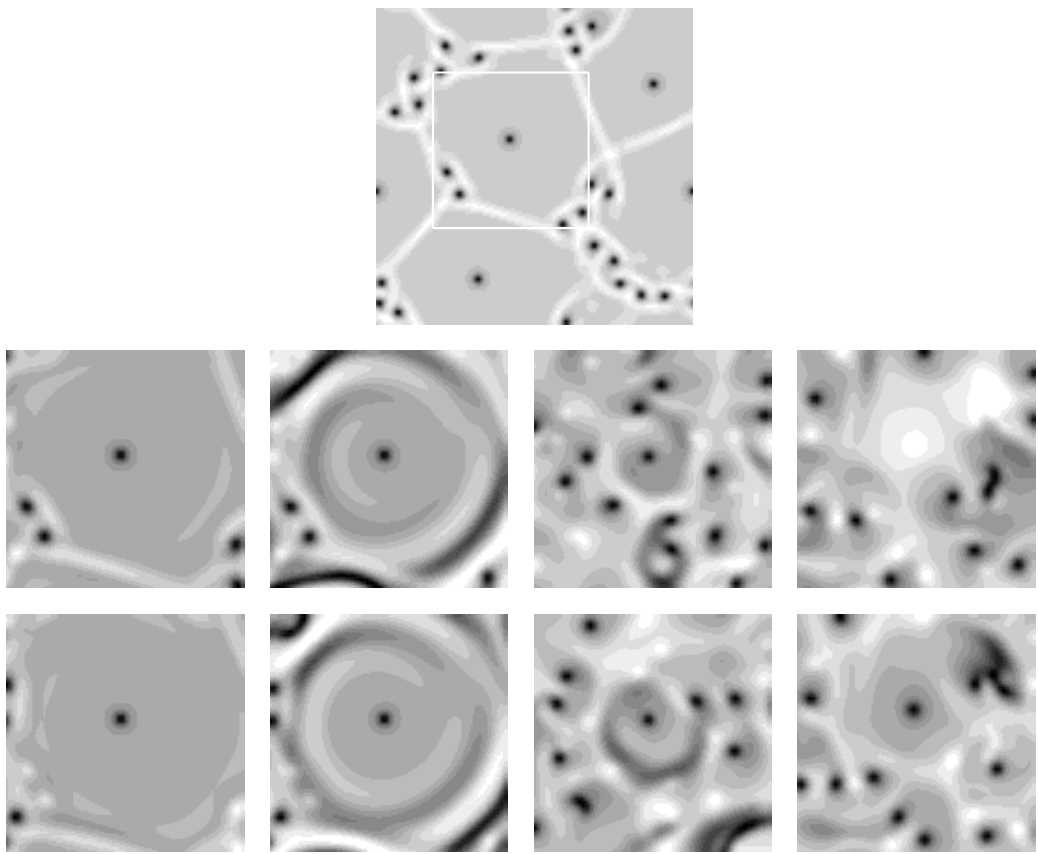,width=14cm} \vspace{1cm}
\caption{Annihilation of a director defect, for $\gamma = 0.35$ and
$\kappa=1.29$. The top figure shows the initial condition
generated with $\gamma=0.1$, the box indicates the domain of the
director defect, and is amplified in the images below. First
row:$|A_+|^2$, second row: $|A_-|^2$. From left to right:
$t=t_0+50$, $t=t_0+100$, $t=t_0+200$, and $t=t_0+300$.  As in the
argument defect case, a scalar defect approach at the end the
defect core, and annihilate one of the components of the director
defect (the one in $A_+$).}
\label{fig:aniq_dir}
\end{figure}

\begin{figure}
\psfig{figure=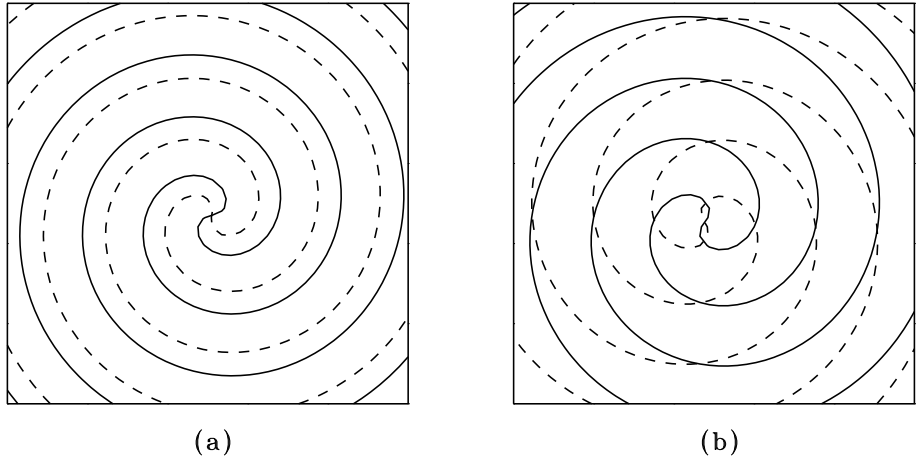,width=14cm} \caption{Lines of constant
phase for (a) an argument defect and (b) a director defect.  Solid
curves correspond to $\phi_+=0$ and $\pi$, and dashed curves to
$\phi_-=0$ and $\pi$.  We can see that in (a) ${\bf k_R} =0$ while
in (b) ${\bf k_R} \neq 0$.} \label{fig:lin_fase}
\end{figure}

\begin{figure}
\psfig{figure=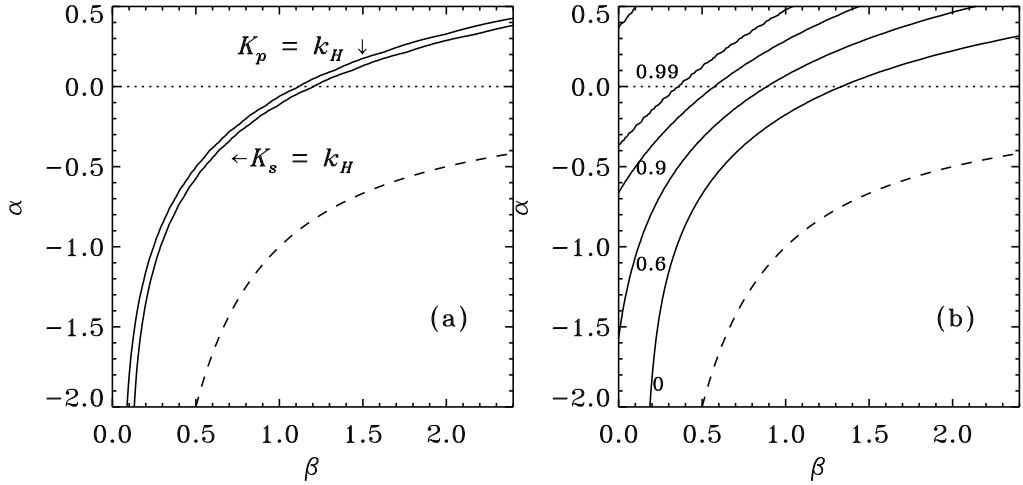,width=14cm} \caption{Stability diagrams in the
$\alpha$-$\beta$ plane showing different stability regions (the
unstable regions are located to the right of the curves).  In both
figures the dashed curve is $1 + \alpha \beta = 0$. In figure (a)
the curves $K_p = k_H$ and $K_s = k_H$ are plotted for $\gamma =
0.3$ (see Eqs. (\protect\ref{kp}) and (\protect\ref{ks})). The
wave emitted by argument defects is unstable to the right of the
$K_p$ curve. The wave emitted by director defects is unstable to
the right of the $K_s$ curve. In figure (b) we plot the curve $K_p
= k_H$ for different values of $\gamma$: 0, 0.6, 0.9 and 0.99 as
indicated.}
\label{fig:ab}
\end{figure}

\begin{figure}
\psfig{figure=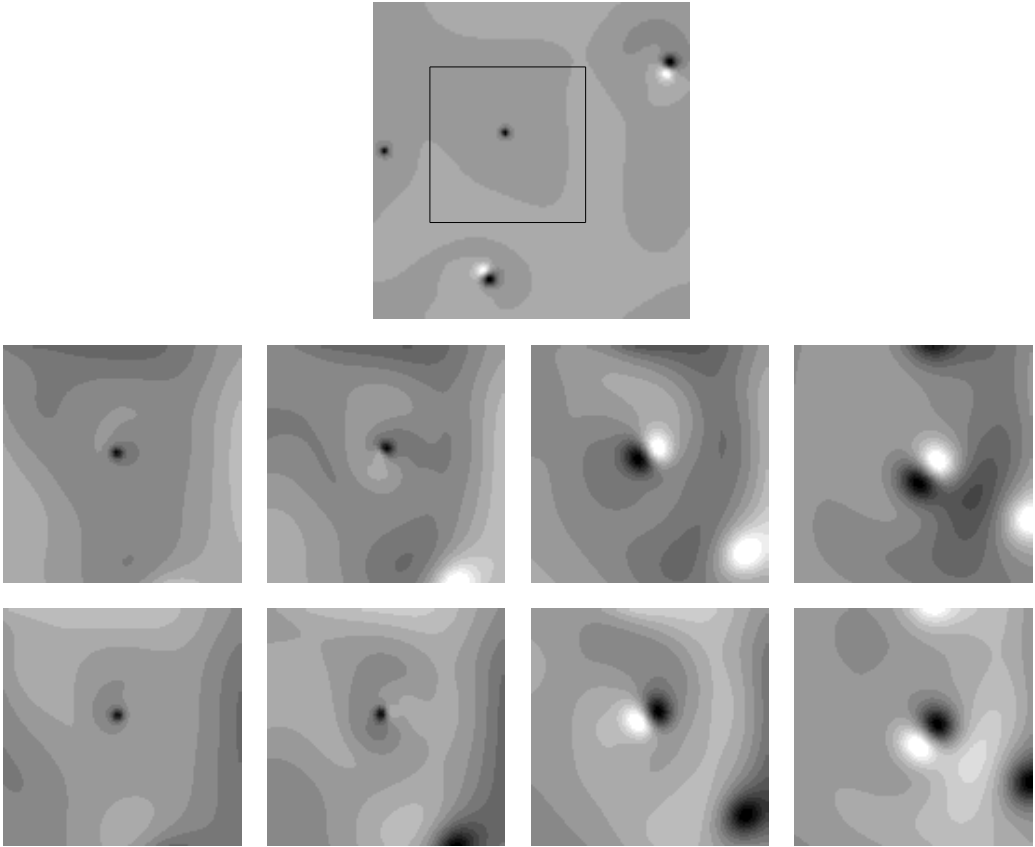,width=14cm} \caption{Splitting of a
director defect, for $\alpha =0.7$, $\beta=2$ ($\kappa=0.54$) and
$\gamma=0.95$. The top figure shows the initial condition
generated with a value of $\gamma=0.9$, the square indicates the
region amplified in the figures below. In the center of the square
there is a director defect. First row: $|A_+|^2$, second row:
$|A_-|^2$. From left to right: $t=t_0+50$, $t=t_0+100$,
$t=t_0+150$, and $t=t_0+200$} \label{fig:split_dir}
\end{figure}

\begin{figure}
\vspace*{8cm} \psfig{figure=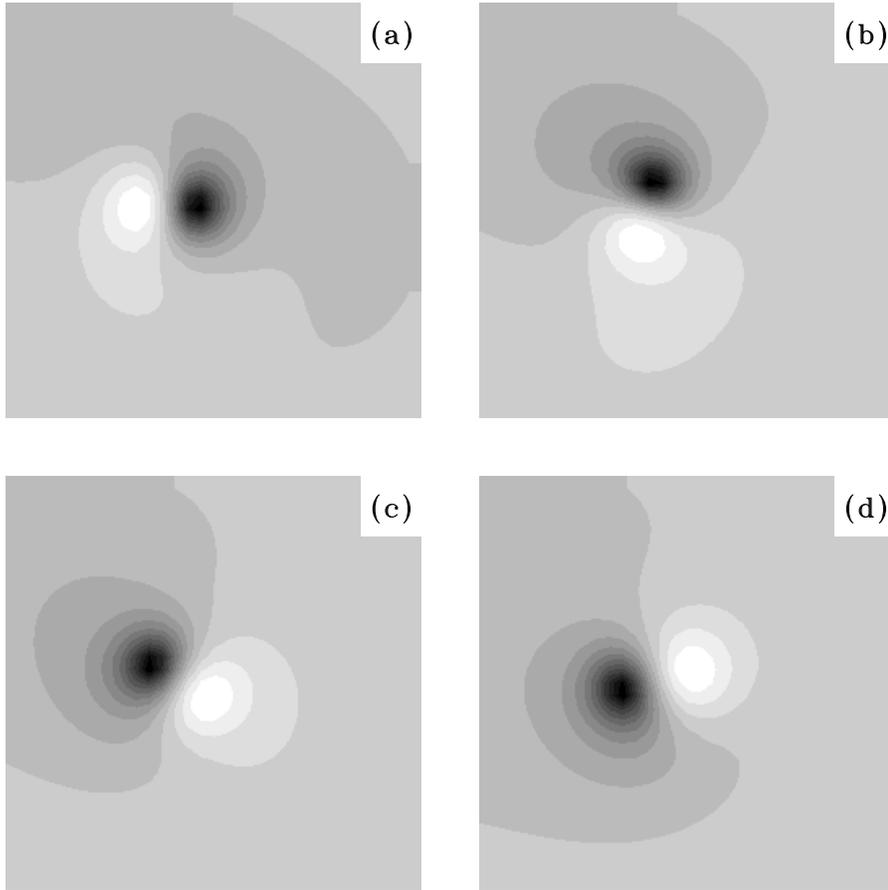,width=14cm} \vspace*{-3cm}
\caption{Time sequence of a bounded pair of scalar defects after
splitting of an argument defect, for $\kappa=1.29$ and
$\gamma=0.9$. The pair rotates and the bounding distance increases
with time.  The quantity plotted is $|A_+|$ for (a) $t=t_0$, (b)
$t=t_0 + 10$, (c) $t=t_0 + 20$, and (b) $t=t_0 + 30$.}
\label{fig:rota}
\end{figure}

\begin{figure}
\psfig{figure=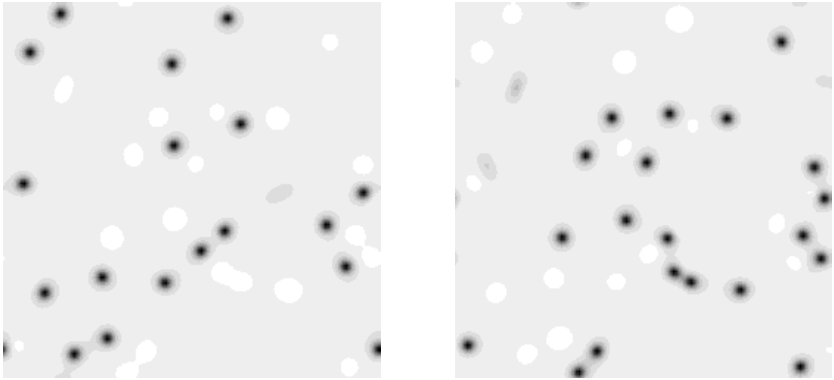} \vspace{1cm} \caption{$|A_+|^2$ and
$|A_-|^2$ for $\alpha = 0$, $\beta = 0$  ($\kappa=0$, potential
limit), $\gamma = 0.1$ and $t=100$ starting from random initial
conditions. There is not spontaneous formation of vectorial
defects for real coefficients.} \label{fig:real}
\end{figure}



\begin{figure}
\psfig{figure=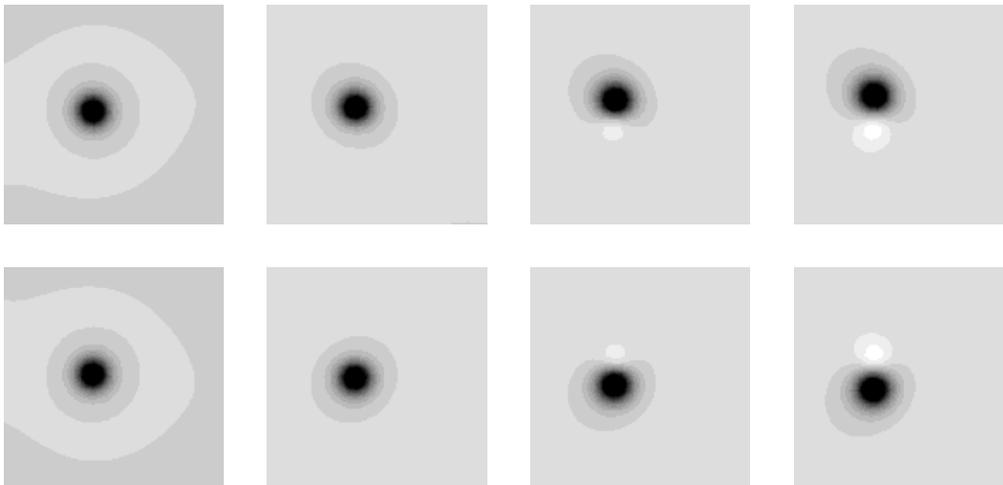, width=14cm} \vspace{1cm} \caption{Temporal
sequence showing the splitting of a vectorial defect for real
coefficients ($\alpha=\beta=0$) and $\gamma = 0.2$. First row:
$|A_+|^2$, second row: $|A_-|^2$.  From left to right $t=50$, 100,
150 and 200.} \label{fig:split}
\end{figure}

\begin{figure}
\psfig{figure=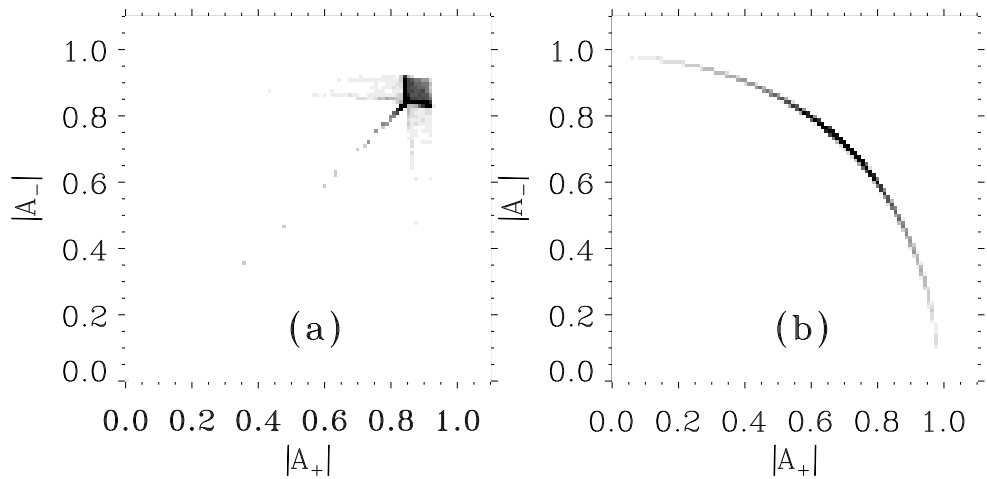,width=14cm} \caption{Joint probability
distribution $p(|A_+|,|A_-|)$ for (a) $\gamma = 0.1$ and (b)
$\gamma = 0.95$ (in both cases $\kappa=1.29$). The histograms that
are presented in grey levels have been constructed by collecting
the values of $(|A_+|,|A_-|)$ at all space points of the
simulation domain, and at $100$ temporal samples separated by
$\Delta t = 1$. The straight diagonal line in the first plot is a
signature of the presence of vector defects ($|A_+|=|A_-|$)
whereas the curve in (b) displays the anticorrelation between
components characteristic of scalar defects. }
\label{fig:joint_prob}
\end{figure}

\begin{figure}
\psfig{figure=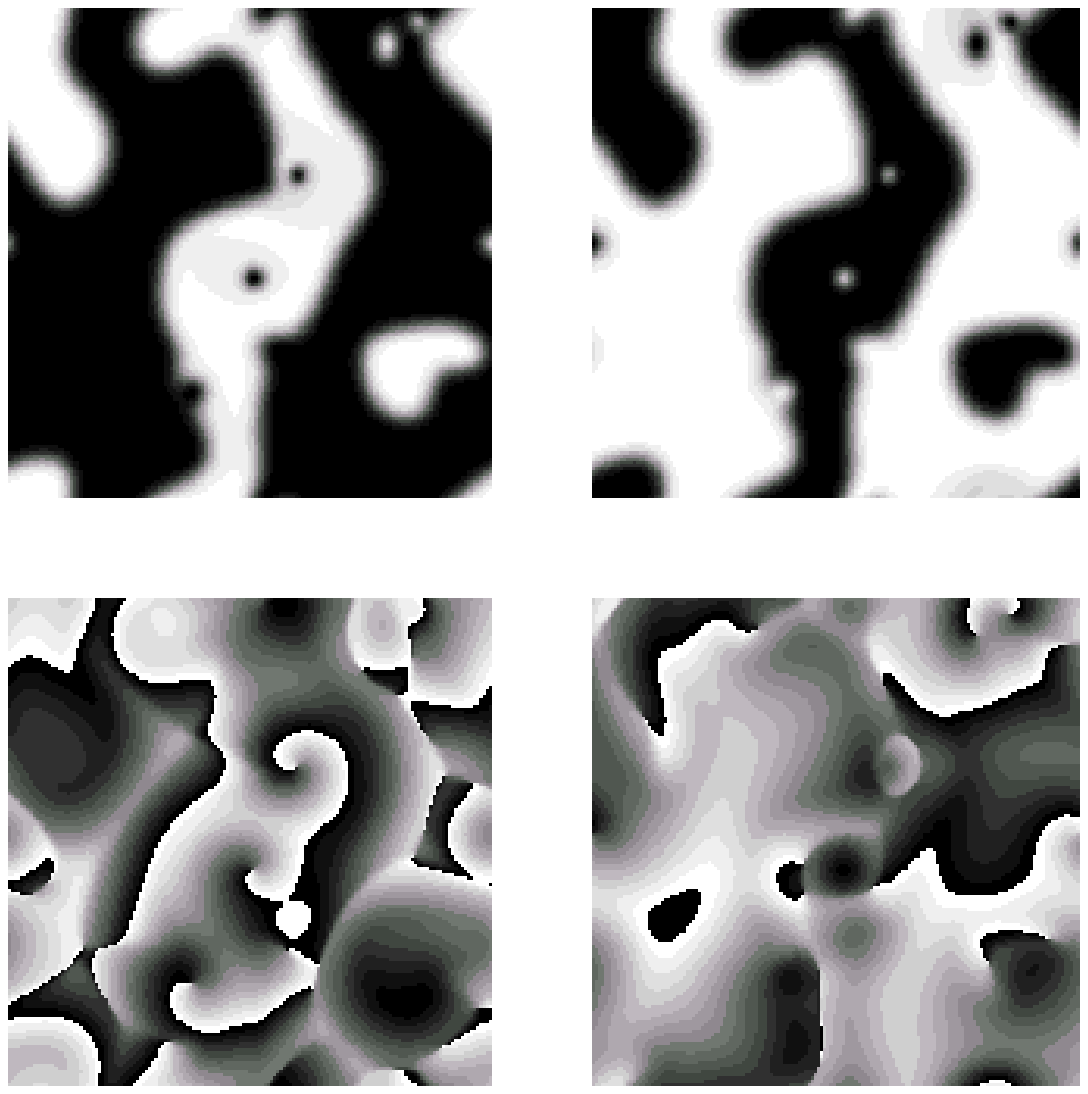,width=14cm} \caption{Configuration of
$|A_+|^2$ and $|A_-|^2$ (top row) and the corresponding phases
(bottom row) for $\gamma = 1.1$, $\alpha=0.2$ and $\beta=2$
($\kappa=1.29$), at $t=50$. The system segregates into two
circularly polarized phases, with some defects (the black dots in
the amplitude plots, associated to the phase singularities in the
phase plots).}
\label{fig:g11}
\end{figure}

\end{center}


\begin{thebibliography}{00}


\bibitem{riecke-pismen} H. Riecke, in {\it Pattern formation in continuous
and coupled systems}, ed. by M. Golubitsky, D. Luss and S. H.
Strogatz (Springer, New York, 1999); L. M. Pismen, {\it Vortices
in nonlinear fields}, (Oxford Univ. Press, New York, 1999).

\bibitem{nonlinearoptics} F.T. Arecchi {\sl et al.}
Phys. Rev. Lett. {\bf 67}, 3751 (1991); K. Staliunas, G. Slekys
and C. O. Weiss, Phys. Rev. Lett. {\bf 79}, 2658 (1997); C. O.
Weiss {\sl et al.} Appl. Phys. {\bf B 68}, 151 (1999);
N. N. Rosanov, in {\em Progress in Optics. Vol. 35},
Ed. by E. Wolf (North-Holland, Amsterdam, 1996);
M. Brambilla {\sl et al.} Phys. Rev. Lett. {\bf 79}, 2042 (1997);
D. Michaelis, U. Peschel and F. Lederer, Phys. Rev. {\bf A 56},
R3366 (1997);
M. Tlidi, P. Mandel and R. Lefever, Phys. Rev. Lett. {\bf 73}, 640
(1994); W. Firth and A.J. Scroggie, Phys. Rev. Lett. {\bf 76},
1623 (1996); C. Etrich, U. Peschel, and F. Lederer, Phys. Rev.
Lett. {\bf 79}, 2454 (1997); K. Staliunas and J.V.
S\'anchez-Morcillo, Phys. Rev. {\bf A 57}, 1454 (1998); G.L. Oppo,
A.J. Scroggie, and W.J. Firth, J. Opt. B: Quant. Semiclass. Opt.
{\bf 1}, 133 (1999).

\bibitem{bhandari} R. Bhandari, Phys. Rep {\bf 281}, 1 (1997).


\bibitem{marco} M. Santagiustina, E. Hern\'{a}ndez-Garc\'{\i}a, M. San Miguel,
A.J. Scroggie, G.-L. Oppo, {\sl Polarisation Patterns and
Vectorial Defects in Type II Optical Parametric Oscillators},
submitted (2001).

\bibitem{rafa} R. Gallego, M. San Miguel and R. Toral, Phys.
Rev. E {\bf 61}, 2241 (2000).

\bibitem{cross} M. C. Cross and P. C. Hohenberg, Rev. Mod. Phys
{\bf 65}, 851 (1993).

\bibitem{bohr} T. Bohr, M. Jensen, G. Paladin, and A. Vulpiani, {\sl
Dynamical Systems Approach to Turbulence} (Cambridge Univ. Press,
Cambridge, 1998).

\bibitem{aranson_kramer} I. S. Aranson, L. Kramer, {\it The World
of Complex Ginzburg-Landau Equation}, preprint (2001).

\bibitem{gil} L. Gil, Phys. Rev. Lett. {\bf 70}, 162 (1993).

\bibitem{gilIJBC} L. Gil, Int. J. Bif. Chaos {\bf 3}, 573 (1993).

\bibitem{pismenPRL} L. M. Pismen, Phys. Rev. Lett. {\bf 72}, 2557,
(1994).

\bibitem{pismen} L. M. Pismen, Physica {\bf D 73}, 244 (1994).

\bibitem{sanmiguel} M. San Miguel, Phys. Rev. Lett. {\bf 75}, 425 (1995).

\bibitem{hernandez} E. Hern\'{a}ndez-Garc\'{\i}a, M. Hoyuelos, P. Colet,
M. San Miguel and R. Montagne, Int. J. Bif. and Chaos {\bf 9},
2257 (1999). 

\bibitem{hoyuelos} M. Hoyuelos, E. Hern\'{a}ndez-Garc\'{\i}a, P. Colet and
M. San Miguel, Comp. Phys. Comm. {\bf 121-122}, 414 (1999).

\bibitem{prl} E. Hern\'{a}ndez-Garc\'{\i}a, M. Hoyuelos, P. Colet, M. San Miguel,
Phys. Rev. Lett. {\bf 85}, 744 (2000).

\bibitem{aranson-pismen} I. S. Aranson and L. M. Pismen, Phys.
Rev. Lett. {\bf 84}, 634 (2000).

\bibitem{amengual} A. Amengual, E. Hern\'andez-Garc\'{\i}a, R. Montagne and M.
San Miguel, Phys. Rev. Lett. {\bf 78}, 4379 (1997).

\bibitem{montagnePL} R. Montagne and E. Hern\'{a}ndez-Garc\'{\i}a, Phys.
Lett. {\bf A 273}, 239 (2000).

\bibitem{potential1} R. Montagne, E. Hern\'{a}ndez-Garc\'{\i}a, and M. San Miguel,
Physica {\bf D 96}, 47 (1996).

\bibitem{potential2} M. San Miguel, R. Montagne, A. Amengual, and
E. Hern\'{a}ndez-Garc\'{\i}a, in {\sl Instabilities and
non-equilibrium structutures V}, edited by  E. Tirapegui and W.
Zeller (Kluwer, Dordrecht, 1996).

\bibitem{bose} A.J. Leggett, Rev. Mod. Phys. {\bf 73}, 307 (2001).

\bibitem{fiber} J. Yang, Y. Tan, Phys. Rev. Lett. {\bf 85}, 3624
(2000), and references therein.

\bibitem{numerics} The integration algorithm is the one presented
in \cite{hoyuelos}, which is a generalization to the vector case
of the pseudospectral in space and second-order in time algorithm
for the CGL equation of \cite{phastur}. We discretize with a $128
\times 128$ square lattice a domain of $64 \times 64$ space units
with periodic boundary conditions. Animations of the different
dynamic phenomena can be downloaded from {\tt
http://www.imedea.uib.es/PhysDept/Nonlinear/research\verb+_+topics/Vcgl2/}.

\bibitem{hagan} P. S. Hagan, SIAM J. Appl. Math. {\bf 42}, 762 (1982).

\bibitem{phastur} R. Montagne, E. Hern\'{a}ndez-Garc\'{\i}a, A. Amengual, and M.
San Miguel, Phys. Rev. {\bf E 56}, 151 (1997).

\bibitem{lega} J. Lega, C. R. Acad. Sci. Paris {\bf 309}(II), 1401 (1989).

\bibitem{victor1} V. M. Egu\'{\i}luz, E. Hern\'{a}ndez-Garc\'{\i}a and O. Piro,
Int. J. of Bif. and Chaos {\bf 9}, 2209 (1999).

\bibitem{victor2} V. M. Eguiluz, E. Hern\'{a}ndez-Garc\'{\i}a and O. Piro,
Phys. Rev. {\bf E 64}, 036205 (2001).

\bibitem{ott_inho} M. Hendrey, K. Nam, P. Guzdar, and E. Ott,
Phys. Rev. {\bf E 62}, 7627 (2000).

\bibitem{vectornote} Earlier theoretical analysis
\cite{pismenPRL} left open the possibility for the existence of
vector defects below a critical value of the coupling, $\gamma_c
\simeq 0.492$, for $\alpha=\beta=0$, but the more recent work
\cite{aranson-pismen} rules out this possibility, or equivalently
$\gamma_c=0$.

\bibitem{aranson} I. S. Aranson, L. Aranson, L. Kramer and A. Weber,
Phys. Rev. A {\bf 46}, R2992 (1992); I. S. Aranson, L. Kramer and
A. Weber, in {\it Instabilities and Nonequilibrium Strutctures
IV}, pag. 259, edited by E. Tirapegui and W. Zeller (Kluwer
Academic Publishers, 1993).

\bibitem{mc} R. J. McEliece, {\it The theory of Information and Coding: a
Mathematical Framework for Communication}, Encyclopedia of
Mathematics and its Applications, Vol. III (Addison-Wesley, New
York, 1977).

\bibitem{gomila} D. Gomila, P. Colet, G.-L. Oppo, M. San Miguel, {\sl
Stable droplets and growth laws close to the modulational
instability of a domain wall}, Phys. Rev. Lett., to appear (2001).


\end{thebibliography}
\end{document}